\documentclass{aa}  
\usepackage{graphicx}
\usepackage{txfonts}
\usepackage{subfig}
\usepackage{color}

\usepackage[normalem]{ulem}

\newcommand{\norm}[1]{\left\lVert#1\right\rVert}

\usepackage{natbib,twoopt}
\usepackage[colorlinks=true, linkcolor=blue, urlcolor=blue, citecolor=blue, breaklinks=true]{hyperref} %% to avoid \citeads line fills
\bibpunct{(}{)}{;}{a}{}{,} %% natbib format for A&A and ApJ
\makeatletter
\newcommandtwoopt{\citeads}[3][][]{\href{http://adsabs.harvard.edu/abs/#3}%
{\def\hyper@linkstart##1##2{}%
\let\hyper@linkend\@empty\citealp[#1][#2]{#3}}}
\newcommandtwoopt{\citepads}[3][][]{\href{http://adsabs.harvard.edu/abs/#3}%
{\def\hyper@linkstart##1##2{}%
\let\hyper@linkend\@empty\citep[#1][#2]{#3}}}
\newcommandtwoopt{\citetads}[3][][]{\href{http://adsabs.harvard.edu/abs/#3}%
{\def\hyper@linkstart##1##2{}%
\let\hyper@linkend\@empty\citet[#1][#2]{#3}}}
\newcommandtwoopt{\citeyearads}[3][][]%
{\href{http://adsabs.harvard.edu/abs/#3}
{\def\hyper@linkstart##1##2{}%
\let\hyper@linkend\@empty\citeyear[#1][#2]{#3}}}
\makeatother

\begin{document} 

   \title{Heating of the solar chromosphere through current dissipation}

   \subtitle{}

   \author{J. M. da Silva Santos\inst{1,2} \and S. Danilovic\inst{1} \and J. Leenaarts\inst{1} \and J. de la Cruz Rodríguez\inst{1} \and X. Zhu\inst{3,4} \and S. M. White\inst{5} \and \\ G.~J.~M.~Vissers\inst{1} \and M. Rempel\inst{6}
          }

   \institute{Institute for Solar Physics, Department of Astronomy, Stockholm University, AlbaNova University Centre, SE-106 91 Stockholm, Sweden 
         \and
            National Solar Observatory, 3665 Discovery Drive, Boulder, CO 80303, USA\\ \email{jdasilvasantos@nso.edu}
         \and
            Key Laboratory of Solar Activity, National Astronomical Observatories, Chinese Academy of Sciences, Beijing 100012, China
         \and 
            Max-Planck-Institut für Sonnensystemforschung, Justus-von-Liebig-Weg 3, 37077 Göttingen, Germany
         \and
            Space Vehicles Directorate, Air Force Research Laboratory, Albuquerque, NM, USA
         \and 
            High Altitude Observatory, National Center for Atmospheric Research, 80307, Boulder, CO, USA
             }

   \date{}

% \abstract{}{}{}{}{} 
% 5 {} token are mandatory
 
  \abstract
   {The solar chromosphere is heated to temperatures higher than predicted by radiative equilibrium. This excess heating is greater in active regions where the magnetic field is stronger.} 
   {We aim to investigate the magnetic topology associated with an area of enhanced millimeter (mm) brightness temperatures in a solar active region mapped by the Atacama Large Millimeter/submillimeter Array (ALMA) using spectropolarimetric co-observations with the 1-m Swedish Solar Telescope (SST).}
   {We used Milne-Eddington inversions, nonlocal thermodynamic equilibrium (non-LTE) inversions, and a magnetohydrostatic extrapolation to obtain constraints on the three-dimensional (3D) stratification of temperature, magnetic field, and radiative energy losses. We compared the observations to a snapshot of a magnetohydrodynamics simulation and investigate the formation of the thermal continuum at 3\,mm using contribution functions.}
   {We find enhanced heating rates in the upper chromosphere of up to $\sim$5\,kW\,m$^{-2}$, where small-scale emerging loops interact with the overlying magnetic canopy leading to current sheets as shown by the magnetic field extrapolation. Our estimates are about a factor of two higher than canonical values, but they are limited by the ALMA spatial resolution ($\sim$\,$1.2^{\prime\prime}$). Band\,3 brightness temperatures reach about $\sim$\,$10^{4}$\,K in the region, and the transverse magnetic field strength inferred from the non-LTE inversions is on the order of $\sim$\,500\,G in the chromosphere.}
   {We are able to quantitatively reproduce many of the observed features including the integrated radiative losses in our numerical simulation. We conclude that the heating is caused by dissipation in current sheets. However, the simulation shows a complex stratification in the flux emergence region where distinct layers may contribute significantly to the emission in the mm continuum.}


   \maketitle
%
% ------------------------------------------------------------------------------ %
%\captionsetup[figure]{font=small}
\section{Introduction}
\label{section:introduction}

The heating of the solar chromosphere and corona is mediated by magnetic fields, which guide Poynting flux generated by plasma motions in the convection zone into the outer atmosphere and provide a means to convert the transported energy into heat. 
The conversion mechanisms can be broadly divided into dissipation of wave energy \citep[e.g.,][]{2020SSRv..216..140V} and dissipation of electric currents induced by the slow evolution of the magnetic field \citep[e.g.,][]{1988ApJ...330..474P}, but their relative contribution to the energy balance is still unclear. In the quiet-Sun (QS) chromosphere, heating by acoustic waves may \citep{2020A&A...642A..52A} or may not \citep{2021ApJ...920..125M} act as the dominant process, but active regions (ARs) definitely require other energy sources.

Recent research has focused mostly on the corona \citep[e.g.,][]{2011Natur.475..477M,2013Natur.493..501C,2015ApJ...811..106H,2015RSPTA.37340256K} but much less on the chromosphere, which requires more sophisticated models that only recently became sufficiently realistic to allow for quantitative comparisons with observations \citep{2019ARA&A..57..189C}. Furthermore, it is not trivial to interpret spectral diagnostics formed under optically thick, nonlocal thermodynamic equilibrium (non-LTE) conditions, such as the resonance lines of \ion{Mg}{II} and \ion{Ca}{II}, which require detailed three-dimensional (3D) radiative transfer calculations including partial frequency redistribution for full accuracy \citep[see a review by][]{2017SSRv..210..109D}. 
However, energy losses are much greater in the chromosphere than in the corona: canonical estimates of average (total) losses for QS and AR are 4\,kW\,m$^{-2}$ and  20\,kW\,m$^{-2}$ for the chromosphere but only 0.3\,kW\,m$^{-2}$ and $<$\,10\,kW\,m$^{-2}$, respectively, for the corona \citep{1977ARA&A..15..363W}. In the era of high-resolution solar physics the focus should shift to obtaining new detailed models that reproduce the observed fine structure rather than averaged observables \citep{2019ARA&A..57..189C}. Yet studies quantifying spatially and temporally resolved energy losses are scarce. Recent estimates with a time resolution of 30\,s and a spatial resolution of $\sim$100\,km indicate losses that can locally be as high as 160\,kW\,m$^{-2}$ in the chromosphere \citep{2020arXiv201206229D}. This was attributed to magnetic reconnection but no attempt was made at reconciling simulations with observations.  

It has been well established that radiative cooling is stronger in regions where the magnetic field is more concentrated, but this relationship is not linear \citep[e.g.,][]{1987A&A...180..241S,1999ApJ...515..812H,2018A&A...619A...5B}. Low resolution ($\gtrsim$\,$10^{\prime\prime}$) observations in the millimeter range show brightness enhancements associated with network and ARs \citep[e.g.,][]{1991ApJ...383..443L,2009A&A...497..273L}, while high-resolution ($\sim$\,$0.1^{\prime\prime}$) optical observations show that the \ion{Ca}{II} K brightness in ARs is dominated by an extended component that is associated with more space-filling, horizontal magnetic fields \citep{2018A&A...612A..28L}. 

Dissipation of electric currents induced by the magnetic field is a prime candidate for explaining this heating, but determining the electric current vector in the chromosphere is notoriously challenging. It has only been reported in sunspots using \ion{Ca}{II} 8542\,\AA~observations \citep{2005ApJ...633L..57S,2021A&A...652L...4L} and in pores using the \ion{He}{I}\,10830\,\AA~multiplet \citep{2003Natur.425..692S}. While \ion{Ca}{II} is partially sensitive to temperatures, the \ion{He}{I} line is not, and thus cannot be used to establish a direct link between heating and electric currents in the atmosphere. 

Observational evidence for magnetic reconnection in the lower atmosphere comes from the association of small-scale optical or ultraviolet (UV) brightenings, such as Ellerman bombs, UV bursts, and jets near or above patches of opposite magnetic polarity in the photosphere \citep[e.g.,][]{2013ApJ...774...32V,2017A&A...605A..49C,2019ApJ...887...56T,2020A&A...633A..58O}, while numerical simulations show how these events could be related to heating in current sheets \citep[e.g.,][]{2015ApJ...799...79N,2017A&A...601A.122D,2019A&A...626A..33H,2020ApJ...891...52S}.

High-resolution observations of the free-free millimeter (mm) continuum are now provided by the Atacama Large Millimeter/sub-millimeter Array \citep[ALMA,][]{2009IEEEP..97.1463W}. This radiation is optically thick in the chromopshere, and the opacity is dominated by electron-proton collisions \citep[see review by][]{2016SSRv..200....1W}. The mm continuum provides strong temperature constraints in inversions of non-LTE lines \citep{2018A&A...620A.124D,2020A&A...634A..56D}, which should allow us to revise estimates of radiative energy losses that can be used to benchmark numerical simulations. 

Small-scale transient {\it mm-bursts} in ARs have also been linked to magnetic reconnection events \citep{2020A&A...643A..41D}, but the lack of chromospheric magnetic field measurements did not enable a definitive conclusion in this regard. Here, we conducted a follow-up study that combines ALMA Band\,3 (100\,GHz or 3\,mm) data with optical spectropolarimetry obtained at the 1-m Swedish Solar Telescope \citep[SST,][]{2003SPIE.4853..341S}, providing further evidence for heating in the upper chromosphere in an AR by current dissipation of up to $\sim$5\,kW\,m$^{-2}$ at the current Band\,3 spatial resolution ($\sim$\,1.2$^{\prime\prime}$). This is supported by a magnetohydrostatic extrapolation and quantitatively reproduced by a 3D radiative-magnetohydrodynamics (r-MHD) simulation.

\section{Observations}
\label{section:observations}

We obtained simultaneous interferometric brightness temperature, $T_{\rm b}$, maps of the 3\,mm continuum with ALMA and spectropolarimetric observations using the CRISP instrument \citep{2008ApJ...689L..69S} at the SST in the Fe\,I 6173\,\AA~and the Ca\,II 8542\,\AA~lines (hereafter $\lambda6173$ and $\lambda8542$) of NOAA AR 12738 on April 13, 2019. We also use ultraviolet (UV) imaging provided by the Atmospheric Imaging Assembly \citep[AIA,][]{2012SoPh..275...17L} and magnetogram data from the Helioseismic and Magnetic Imager \citep[HMI,][]{2012SoPh..275..207S} on board the Solar Dynamics Observatory \citep[SDO,][]{2012SoPh..275....3P}.

The ALMA data set consists of $T_{\rm b}$ maps taken approximately between 18:20-18:55\,UTC and 19:16-19:51\,UTC at 2\,s cadence with 140\,s calibration intervals every 10\,min. These data have been previously presented by \citet{2020A&A...643A..41D} and we refer to their study for further details of the data reduction and calibration.
In this paper, we only use ALMA maps taken within the time span of the SST observations (see below). The field of view (FOV) has a diameter of $60^{\prime\prime}$ and the pixel scale is $0.3^{\prime\prime}$. The noise root-mean-square level is approximately 20\,K. 

Coordination between ALMA in Chile and the SST in La Palma is challenging because of the difference in time zones. The SST/CRISP observations started as soon as the seeing conditions improved in the late afternoon but it was technically unfeasible to prolong the campaign for a long period given the closeness to local sunset. Therefore, the CRISP spectropolarimetric data (full-Stokes) consist of a single line scan of sufficient quality in 17 wavelength positions in $\lambda8542$ in the range of $\pm700$\,m\AA~and 15 positions in $\lambda6173$ within $\pm 275$\,m\AA~from the line center taken between 18:48:36\,UTC and 18:48:56\,UTC. 

\begin{figure}[t]
    \centering
    \includegraphics[width=\linewidth]{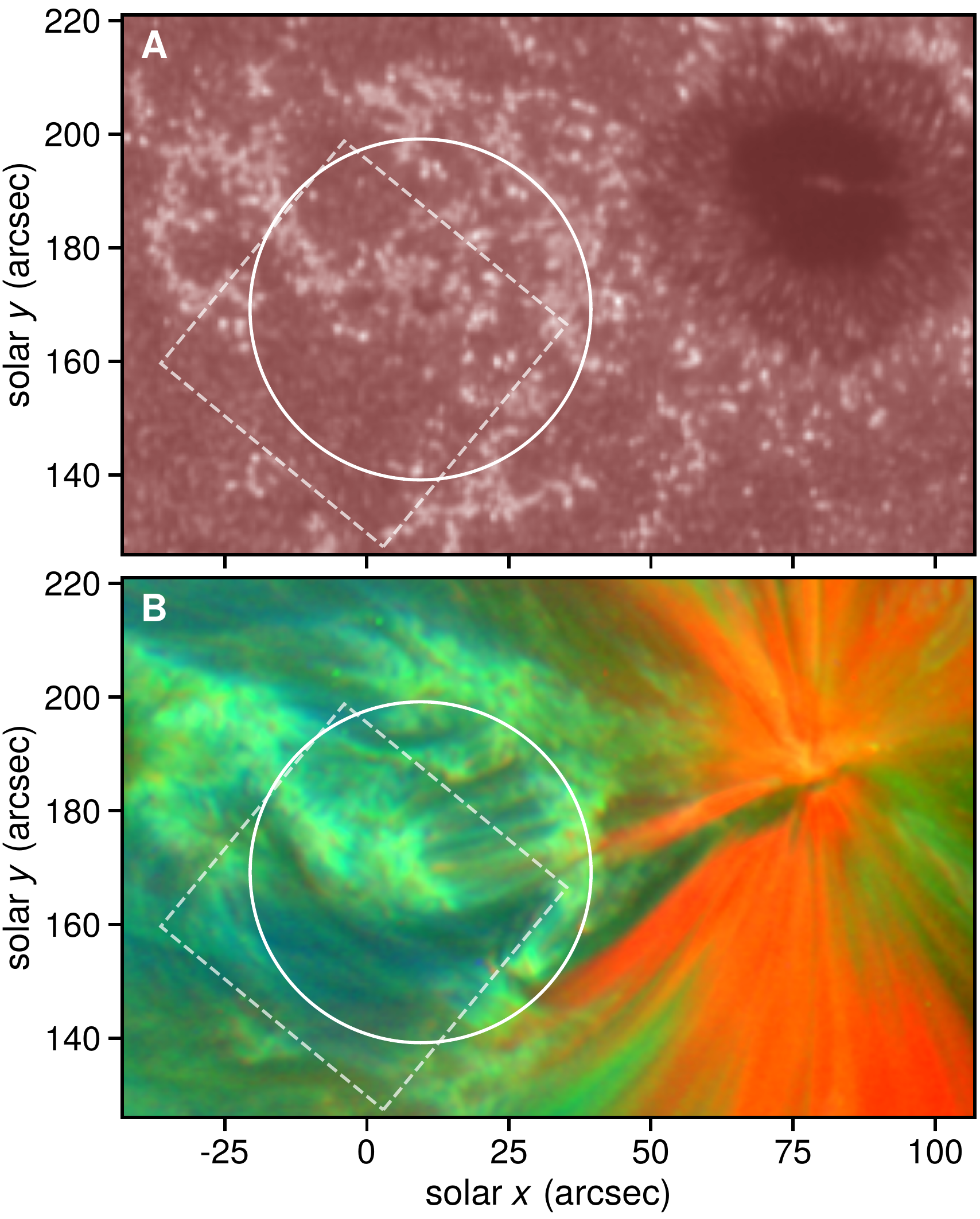}
    \caption{Extended view of NOAA AR 12738 provided by SDO. Panel A: SDO/AIA\,1700\,\AA~intensity in square-root scale; panel B: Composite of AIA 171\,\AA~(red), 193\,\AA~(green), and 211\,\AA~(blue) intensities (unsharpened). Solid circle and dashed square show the ALMA and SST fields, respectively.} 
    \label{fig:Spl1}
\end{figure}

\begin{figure*}[t]
    \centering
    \includegraphics[width=\textwidth]{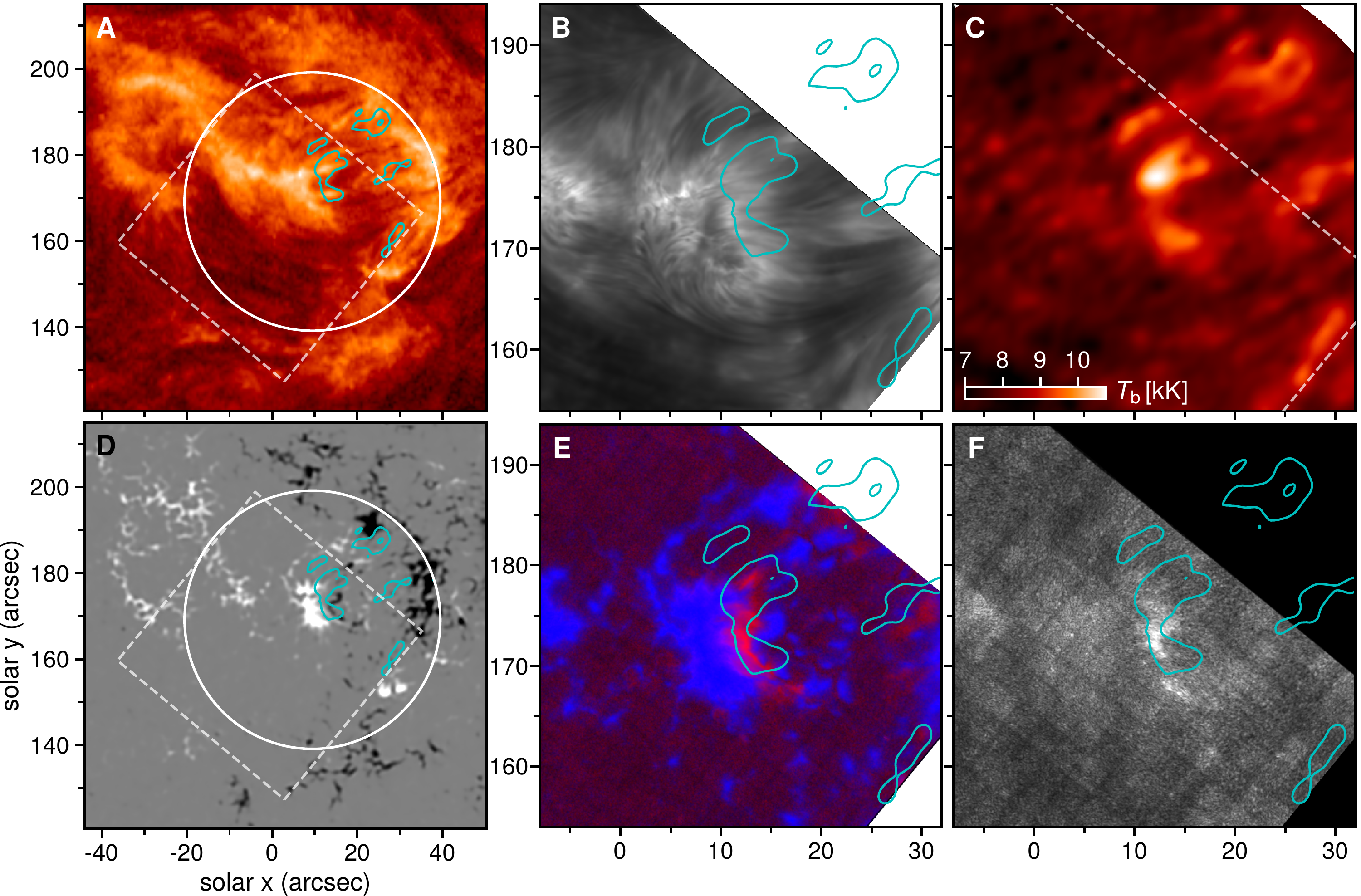}
    \caption{Multiwavelength imaging of a solar active region. The leftmost panels show an extended view of the target, while the ones on the right show a closer look at the center. Panel A: AIA\,304\,\AA~intensity in logarithmic scale (18:48:53\,UTC); panel B: SST/CRISP $\lambda$8542\,\AA~core (18:48:41\,UTC); panel C: ALMA brightness temperature at 3\,mm (18:48:53\,UTC); panel D: HMI LOS magnetogram clipped at +/- 1\,kG (white/black, 18:49:30\,UTC); panel E: Composite of total linear polarization (red) and total circular polarization (blue) in $\lambda$6173 (18:48:53\,UTC); panel F: TLP in $\lambda$8542 (18:48:41\,UTC). Solid circle and dashed square show the ALMA and SST fields. The cyan contours correspond to $T_{\mathrm{b}}[3\,\mathrm{mm}] = 9$\,kK.}
    \label{fig:fig1}
\end{figure*}

The SST data were reduced using the \texttt{CRISPRED} pipeline \citep{2015A&A...573A..40D}, which includes flat-field and dark correction, cross-talk correction, polarimetric calibration, image reconstruction through Multi-Object Multi-Frame Blind Deconvolution \citep[MOMFBD,][]{2002SPIE.4792..146L,2005SoPh..228..191V}, and fringe removal using Fourier filtering. An additional fringe removal step using principal component analysis was necessary in order to remove large-scale patterns in Stokes Q and U \citep{2020A&A...644A..43P}. 
The absolute intensity and wavelength calibrations were performed using the solar atlas of \citet{1984SoPh...90..205N} as reference. The pixel scale is $0.059^{\prime\prime}$. 
We co-aligned the CRISP and HMI data by cross-correlating the 6173\,\AA~continuum images taken by both instruments. The calibrated SST and ALMA data sets were analyzed using inversion codes (Section\,\ref{section:inv}).

The UV and extreme-UV (EUV) images taken by AIA and the line-of-sight (LOS) magnetograms obtained by HMI were downloaded from the Virtual Solar Observatory using the \texttt{vso\_search} routine in SolarSoftWare (SSW) \citep{1998SoPh..182..497F}; they were further processed using IDL tools in SSW and the \texttt{SunPy} package \citep{2015CS&D....8a4009S} as described in \citet{2020A&A...643A..41D}.  
The LOS magnetograms were deconvolved using \texttt{Enhance}\footnote{\url{https://github.com/cdiazbas/enhance}} \citep{2018A&A...614A...5D}.
In addition, we used the full-vector magnetogram that was closest in time (taken at 18:48:01\,UTC) to the CRISP scans for the magnetic field extrapolation (Section\,\ref{section:extrapol}). 
The vector magnetogram has been processed with the \texttt{SHARP} pipeline \citep{2014SoPh..289.3549B} and it was obtained from the JSOC interface \footnote{\url{http://jsoc.stanford.edu/HMI/HARPS.html}}.

Figure\,\ref{fig:Spl1} shows a context view of the AR provided by SDO. The SST and ALMA pointings were on a group of pores and an arch-filament system (AFS) southwest of a large sunspot close to disk center. The EUV composite image was generated using the \texttt{make\_lupton\linebreak\_rgb} function in \texttt{Astropy} \citep{astropy:2018,2004PASP..116..133L}. The focus of this paper is the area surrounding the pore at the center of the ALMA FOV, which is the location of the western footpoints of the AFS.

\section{Methods}
\label{section:methods}

\subsection{Data inversions}
\label{section:inv}

We ran two different inversion codes on the SST data for different purposes: \texttt{PyMilne}\footnote{\url{https://github.com/jaimedelacruz/pyMilne}} \citep{2019A&A...631A.153D}, a code based on the Milne-Eddington (ME) approximation for photospheric lines using analytic response functions \citep{2007A&A...462.1137O}; and \texttt{STiC}\footnote{\url{https://github.com/jaimedelacruz/stic}} \citep{2019A&A...623A..74D}, a multi-atom, non-LTE inversion code based on the Rybicki-Hummer code \citep[RH,][]{2001ApJ...557..389U} that is suitable for both photospheric and chromospheric lines. The former provides the mean magnetic field vector within the formation region of the lines in a large FOV in a quick manner, whereas the latter requires more computing power but it allows for a detailed investigation of the thermodynamic stratification of the plasma as function of logarithmic optical depth of the 500\,nm continuum (here simply $\log \tau$), while taking radiative transfer effects into account \citep[e.g.,][]{2017SSRv..210..109D}.

\subsubsection{Milne-Eddington inversions}
\label{section:ME}

We fitted the $\lambda$6173 spectra taken by SST/CRISP using \texttt{PyMilne} in order to provide an input for the magnetic field extrapolation code (Section\,\ref{section:extrapol}). We took into account the spectral point-spread-function of the instrument and cavity errors \citep{2015A&A...573A..40D}. The magnetic filing factor is assumed to be unity. The results are shown in the supplementary Fig.\,\ref{fig:ME}.

\begin{figure*}[t]
    \centering
    \sidecaption
    \includegraphics[width=120mm]{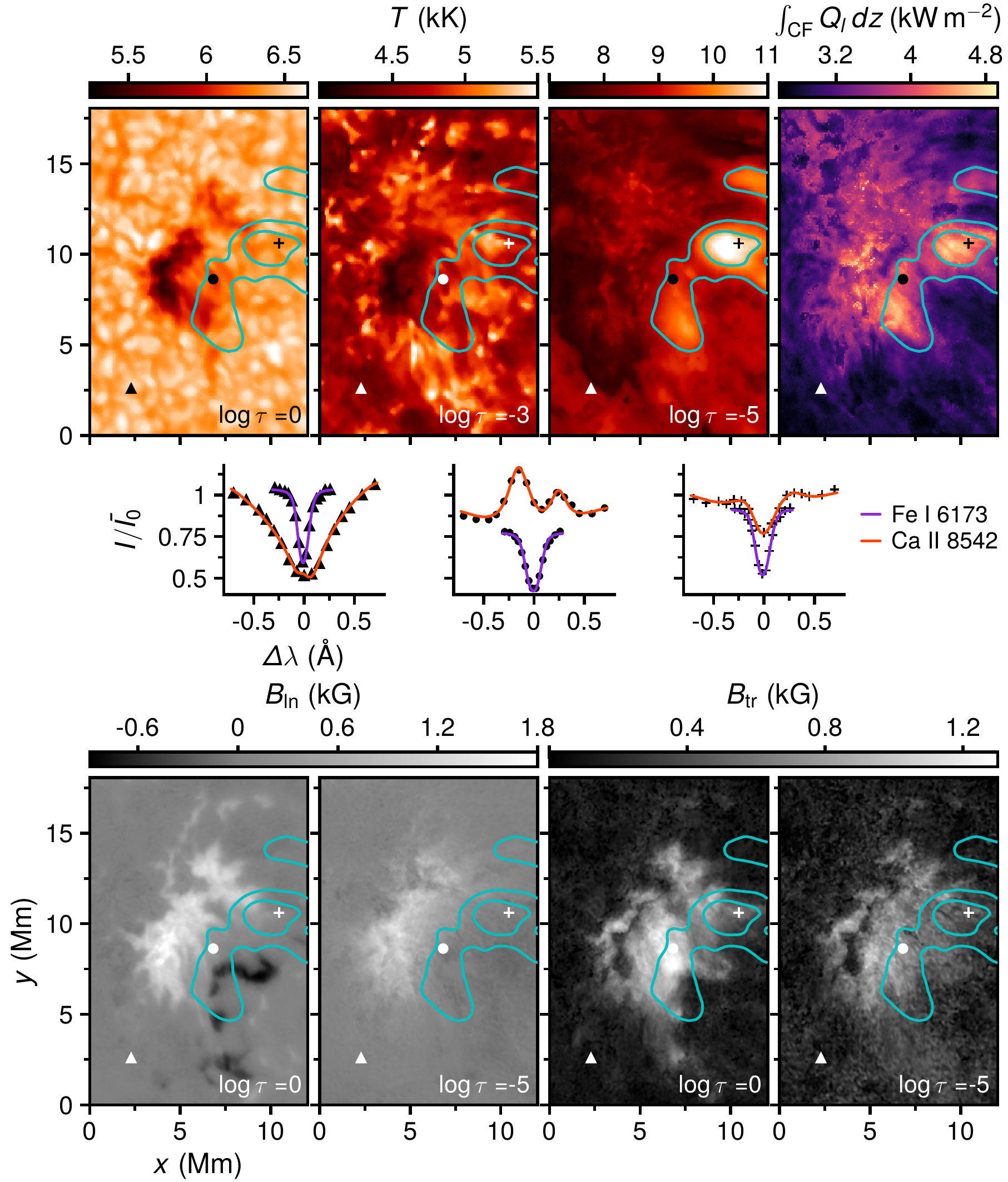}
    \caption{ Non-LTE inversions of the spectral data and radiative losses in the upper chromosphere. Temperature, integrated radiative losses within the contribution function of the 3\,mm continuum, longitudinal field strength, and transverse field strength at selected optical depths from the photosphere to the chromosphere are shown. The cyan contours correspond to $T_{\mathrm{b}}[3\,\mathrm{mm}] = 9$ and 10\,kK. The middle row shows example observed (markers) and fitted (solid lines) intensity of the $\lambda6173$ and $\lambda8542$ lines at three locations indicated by different markers overlaid on the other panels; the intensity is normalized by $\bar{I_{0}}$ -- the mean intensity at the bluest sampled wavelength of each line.}
    \label{fig:fig2}
\end{figure*}

Upper limits on the uncertainty in the parameters can be obtained from their spatial variation on scales shorter than those of typical photospheric features (e.g., granule size $\sim$1$^{\prime\prime}$). Using 5$\times$5\,px boxes at five different locations in the magnetic areas of the region of interest (ROI; see Fig.\,\ref{fig:fig2}), we found average standard deviations of $\delta\norm{B}=27$\,G, $\delta v_{\rm LOS}=0.07\,\rm km\,s^{-1}$, $\delta \theta=2^{\circ}$, and $\delta \phi=8^{\circ}$, in the magnetic field strength, line-of-sight velocity, inclination angle, and azimuth angle, respectively.

\subsubsection{Non-LTE inversions}
\label{section:nlte}

We also ran non-LTE inversions of the $\lambda$6173 and $\lambda$8542 lines along with the ALMA $T_{\rm b}[\rm 3\,mm]$ map using the \texttt{STiC} code similarly to the approach described in \citet{2018A&A...620A.124D}.
The Band\,3 $T_{\rm b}$ maps were converted into intensity in c.g.s. units using the Planck function and linearly resampled to the pixel scale of the CRISP data. The inherent inversion uncertainties (see below) outweigh the interpolation uncertainties.
We note that there is a factor of ten difference in spatial resolution between the optical/infrared and mm diagnostics. Multiresolution spectral data can be dealt with using linear operators and global inversion schemes \citep[e.g.,][]{2019A&A...631A.153D} but this has not yet been implemented into \texttt{STiC}. Therefore, we have decided to preserve the high-resolution information provided by the SST/CRISP spectropolarimetry and oversample the ALMA maps. This approach worked well here because of the distinct formation heights of $\lambda$8542 and the 1.25mm continuum. The 
high-resolution information in the $\lambda$6173 and $\lambda$8542 lines provides the temperature stratification  at high resolution in the lower atmosphere. Because the chromosphere at depths lower than $\log \tau \sim-5$ is so unconstrained from inversions of those spectral lines \citep{2018A&A...620A.124D}, \texttt{STiC} will essentially provide a low-resolution upper chromosphere constrained by ALMA on top of the high-resolution lower atmosphere constrained by the spectral lines.
We restricted the inversions to a $\sim$$16.5^{\prime\prime}$$\times$$24.8^{\prime\prime}$ subfield that encloses the ROI in order to reduce the computational cost. This analysis step required a few million core-hours.

We treated each pixel independently (1.5-D approximation) by simultaneously solving the statistical equilibrium equation for non-LTE populations of a 4-level H atom, a 6-level Ca\,II atom, and a 16-level Fe\,I atom along with the charge conservation equation, while other atoms and molecules are treated in LTE. Compared to LTE, treating the hydrogen atom in non-LTE and correcting the electron densities using charge conservation provides more realistic values that directly impact the calculation of the opacity in the mm continuum \citep[e.g.,][]{2020A&A...634A..56D}. 
We find that treating the Fe\,I atom in non-LTE leads to average differences in the temperature and magnetic field strength of the order of a few percent at $\log \tau=0$ (relative to LTE). Differences in the field strength can be up to $\sim$\,60\% in some patches. \citet{2020A&A...633A.157S} investigated this in detail for the \ion{Fe}{I} 6301, 6302\,\AA~lines using a more complete atomic model, but this would further increase the computational cost of our inversions. It is not critical for our conclusions as we are mainly interested in estimating radiative losses in the chromosphere, however, these effects contribute to the uncertainties of the ME inversions and field extrapolations (Section\,\ref{section:extrapol}).
We then ran another spectral synthesis solving the statistical equilibrium equation for a 11-level Mg\,II atom in order to obtain population densities and radiative rates for computing radiative energy losses (Section\,\ref{section:enlosses}). The calculations include PRD in the \ion{Mg}{II} h and k lines. The gas pressure is obtained by integration of the hydrostatic equilibrium equation and the value at the top boundary is treated as a free parameter \citep{2019A&A...623A..74D}. 

The polarization signals are assumed to be dominated by the Zeeman effect and we do not take the Hanle effect into account. This is a good approximation since the ROI features magnetic fields that are stronger than 500\,G \citep{2021arXiv210604478C}.

\texttt{STiC} uses parameterization by nodes that are interpolated using Bézier splines in an optical depth grid. We used nine nodes in temperature, four nodes in line-of-sight velocity, and two nodes in microturbulence, parallel and transverse magnetic field, and magnetic azimuth angle.

The uncertainties were estimated using a Monte Carlo approach \citep{1992nrca.book.....P} on selected pixels in the ROI (supplementary Fig.\,\ref{fig:STiC_spl}). After obtaining good fits to the SST and ALMA observations, we generated up to 100 different synthetic spectra adding a random component that is the sum in quadrature of white noise and flux calibration uncertainties. The latter amounts to less than 1\% for the spectral lines \citep{1984SoPh...90..205N} and about 5\% for the 3\,mm continua. The synthetic spectra were inverted using different randomly generated atmospheres as initial guesses. This provides probability distributions for each parameter from which we computed the 16th, 50th and 84th percentiles at every optical depth (supplementary Fig.\,\ref{fig:STiC_spl}). 
Typical uncertainties in temperature are of the order of $\sim$50\,K or lower within $\log\tau=[-4,0]$, but they increase at lower optical depths reaching $\sim$600\,K at $\log\tau=-6$; below these values, the response functions of the diagnostics are weak. The uncertainty in the magnetic field strength is $\sim$10-20\,G for the longitudinal (or LOS) component and $\sim$10-70\,G for the transverse component within $\log\tau=[-5,0]$. The sensitivity of the $\lambda$6173 and $\lambda$8542 lines to magnetic fields is weak at optical depths lower than $\log\tau=-5$.

\subsection{Magnetic field extrapolation}
\label{section:extrapol}

We performed a magnetic field extrapolation using a magnetohydrostatic (MHS) model based on the SST and HMI data. Since the SST/CRISP FOV covers only part of the magnetic connectivity, we embedded a cutout of the FOV in the SHARP-processed (disambiguated) HMI vector magnetogram, which covers the entire AR (top panels in supplementary Fig.\,\ref{fig:extrapolSpl}). 
In this way we make use of the high-resolution information provided by the SST data in the ROI and we take advantage of the more extended HMI FOV for context. This is necessary to interpret the interaction of small- and large-scale loop systems. The resulting magnetogram is in cylindrical-equal-area projection (CEA) coordinates.

To ensure consistency between the CRISP and HMI magnetograms, we disambiguated the azimuth angle provided by the ME inversion of the CRISP data by imposing an acute angle with the HMI/SHARP magnetogram. Both magnetograms are derived from the same spectral line ($\lambda6173$). The borders of the CRISP cutout were smoothed with a Gaussian kernel.  
Finally, the maps of the three components of the magnetic field vector, $\vec{B}$, were convolved with a median filter to reduce the impact of bad pixels and inversion noise in the extrapolation. 

The MHS model takes into account the effect of plasma forces which cannot be ignored in the lower atmosphere. An optimization approach is used to numerically solve the equations \citep{2019A&A...631A.162Z}:
\begin{eqnarray}
    \frac{1}{\mu_{0}}(\nabla \times \vec{B}) \times \vec{B} - \nabla p + \rho\,\vec{g} = 0,\\
    \nabla \cdot \vec{B} = 0,
\end{eqnarray}
starting from a nonlinear force-free field extrapolation as initial guesses 
\citep{2004SoPh..219...87W}.
Here, $\mu_0$ is the magnetic permeability, $p$ is the gas pressure, $\rho$ is the mass density, and $\vec{g}$ is the gravitational acceleration. The optimization procedure aims to reach a static equilibrium state between the Lorentz force, the pressure gradient, and gravity.

The model uses the photospheric vector magnetogram (Section\,\ref{section:ME}) as the lower boundary condition. The plasma pressure at the lower boundary is calculated from:
\begin{equation}
    p+\frac{B^2_{z}}{2} = P_\mathrm{ph},
\end{equation}
where $P_\mathrm{ph},$ is a typical photospheric pressure. Further details of the MHS method can be found in \citet{2018ApJ...866..130Z,2020A&A...640A.103Z}.
The resulting model has a size of $2000\times1800\times256$ pixels with a uniform pixel scale of 40\,km.

\subsection{3D radiative-magnetohydrodynamics simulation}
\label{section:simulation}

The simulation presented here was performed with the \texttt{MURaM} code \citep{2005A&A...429..335V,2017ApJ...834...10R}, which includes the following physics: single fluid MHD, 3D grey LTE radiative transfer, a tabulated LTE equation of state, Spitzer heat conduction, and optically thin radiative losses in the corona based on \texttt{CHIANTI} \citep{2012ApJ...744...99L}. The chromospheric and coronal parts of the simulation domain is heated by the Poynting flux generated through magnetoconvection in the photosphere and convection zone. Earlier simulations with this code have reproduced many observed chromospheric phenomena \citep[e.g.,][]{2019A&A...631A..33B,2020LRSP...17....3L}.

The simulation domain has an extent of $40\times 40 \times 22$\,Mm, spanning the vertical direction from $-8$\,Mm to 14\,Mm above the average $\tau_{500\,\mathrm{nm}}$\,$=$\,1 height. The pixel size is 0.078\,Mm. The run was initialized with a bipolar uniform magnetic field of 200\,G \citep{2011A&A...533A..86C}, which is added to the well-developed nonmagnetic convection simulation to form extended magnetic field concentrations at meso- to super-granular scales. This was done with the aim of reproducing the plasma dynamics of solar plage -- regions of moderate magnetic activity. The computational domain was then extended to include the upper solar atmosphere, and the magnetic field from the preexisting simulation was used for potential field extrapolation into the rest of the domain. The new simulation was then run until a relaxed state was achieved. An additional bipolar flux system was advected through the bottom boundary through an ellipsoidal region with major axes (a, b) = (3, 1) Mm and a field strength of 8000\,G \citep{2019NatAs...3..160C}, which emerged from the convection zone into the chromosphere directly beneath the preexisting filaments.  Once the flux reaches the photosphere, its field strength decreased to around 1.5\,kG.

Since the non-LTE spectral synthesis is quite computationally intensive and memory-demanding, here we analyzed one single simulation snapshot where we identified a region of enhanced $T_{\rm b}[3\rm\,mm]$ at a later stage of the flux emergence $t=17$\,min after the magnetic flux reached the photosphere. Anyhow, the lack of spectropolarimetric time series in $\lambda6173$ and $\lambda8542$ does not allow us to compare the observations and the unraveling of the simulated flux emergence. 

We computed the emerging intensities in $\lambda$8542 from the simulation in non-LTE 1.5D (full-Stokes) using \texttt{STiC} and in 3D (intensity only) using \texttt{Multi3D} \citep{2009ASPC..415...87L}. The synthetic \ion{Ca}{II} lines were convolved with the CRISP response function.
The 3\,mm continuum intensities were computed assuming statistical equilibrium and charge conservation using \texttt{STiC} \citep[see also][]{2020A&A...634A..56D}.
Besides the synthetic intensities, we also stored the opacities $\alpha (\nu, z)$ as a function of frequency, $\nu$, and height, $z$ for each pixel. We investigated the formation of the 3\,mm continuum in detail using contribution functions (CF) defined as
\begin{equation}
    \mathrm{CF}_{\nu} (z)=\alpha_{\nu}(z)\,S_{\nu}(z)\,e^{-\tau_{\nu}(z)},
\end{equation}
\noindent where $\nu=100$\,GHz, $S_{\nu}$ is the source function, which is given by the Planck function, and $\tau_{\nu}$ is the optical depth. This quantity essentially quantifies the contribution from different heights of the atmosphere to the emerging intensities.

\subsection{Radiative energy losses}
\label{section:enlosses}

Besides the radiative losses, $Q_{\rm l}$, that can be directly retrieved from the \texttt{MURaM} simulation output \citep{2017ApJ...834...10R}, we also obtained $Q_{\rm l}[\rm STiC]$ from the non-LTE synthesis with \texttt{STiC} by summing the contributions to the radiative cooling from the Ca\,II H, K, infrared triplet, Mg\,II h, k, UV triplet, H$\alpha$, and Ly$\alpha$ lines as follows:
\begin{equation}
    Q_{\rm l}[\mathrm{STiC}] = h\nu_{0}(n_{\rm u}R_{\rm ul} - n_{\rm l}R_{\rm lu}),
\end{equation}
\noindent where $h$ is the Planck constant, $\nu_{0}$ is the frequency of the transition, $n_{\rm u}$ and $n_{\rm l}$ are the population densities of the upper and lower levels, and $R_{\rm ul}$ and $R_{\rm lu}$ are the radiative rates of the transitions
\citep[e.g.,][]{2020arXiv201206229D}. These chromospheric losses are more suitable for comparison with the losses that can be inferred from observations using data inversions (Section\,\ref{section:nlte}).

Total radiative losses were obtained by integration in the height range spanned by the CF of the 3\,mm continuum using a threshold of 1\% of the maximum in each pixel. This criterion was used both for the CFs obtained from the simulation and observations. 

\begin{figure}
    \centering
    \includegraphics[width=\linewidth]{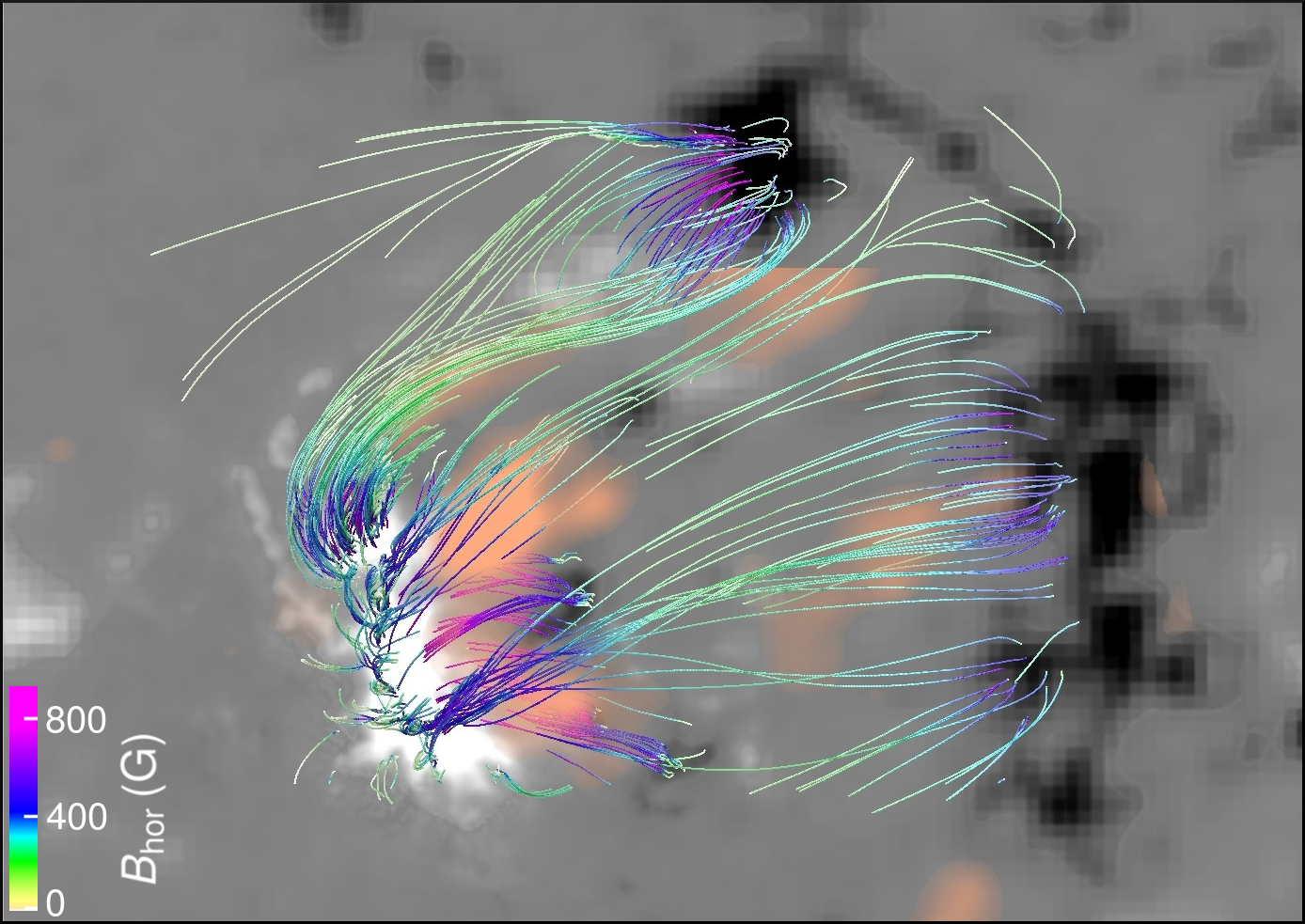}
    \caption{Magnetohydrostatic extrapolation of the SST and HMI composite magnetogram and millimeter continuum brightness. The background shows a SST/CRISP and HMI magnetogram composite image clipped at +/-1\,kG (white/black). The magnetic field lines are computed from the MHS extrapolation and they are color-coded with the horizontal field strength. The pink shade shows regions where $T_{\rm b}\rm\,[3\,mm]>9$\,kK.} 
    \label{fig:fig3}
\end{figure}

\begin{figure*}
    \centering
    \includegraphics[width=\linewidth]{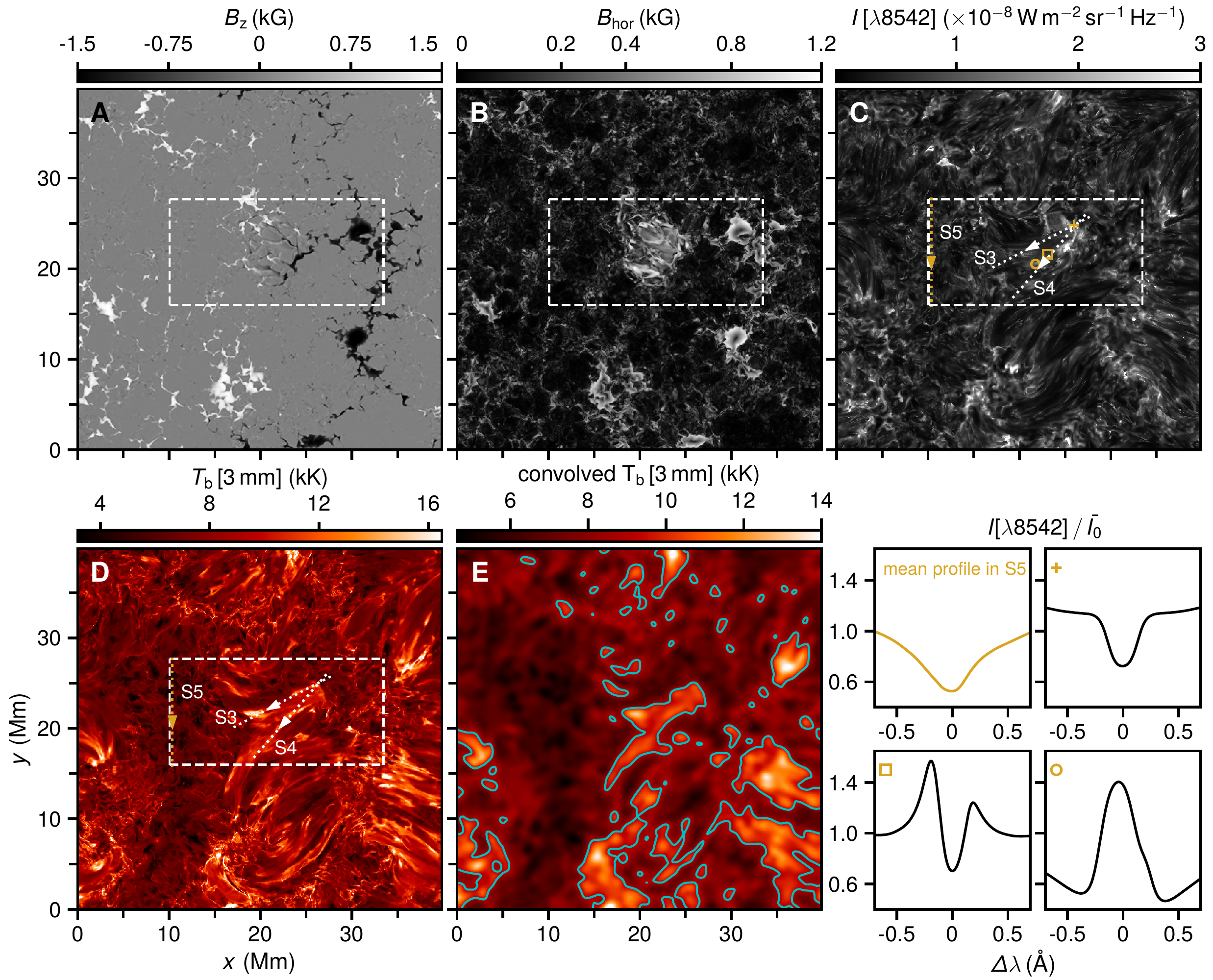}
    \caption{Simulated magnetograms and synthetic emission. Strength of the vertical (panel A) and horizontal (panel B) components of the magnetic field at $z=0$\,Mm; panel B has been gamma-adjusted for display purposes. Panel C: Intensity in the core of $\lambda 8542$; the range is capped for display purposes. Panel D and panel E: Continuum $T_{\rm b}[\rm 3\,mm]$ at full resolution and convolved with a Gaussian kernel with full-width-at-half-maximum of 1.2$^{\prime\prime}$; the cyan contours show $T_{\rm b}[\rm 3\,mm]=9$\,kK. The dashed box delimits the area displayed in Fig.\,\ref{fig:fig4}\textcolor{blue}{B}, \ref{fig:fig4}\textcolor{blue}{C}. The lower right panels show selected $\lambda 8542$ profiles (normalized intensity) in the flux emergence region. Vertical cuts through various parameters of the simulated atmosphere along the slices {\it S3}, {\it S4}, and {\it S5} are displayed in the supplementary Fig.\,\ref{fig:simSlices}.}
    \label{fig:simextra0} 
\end{figure*}

\begin{figure*}[t]
    \centering
    \includegraphics[width=0.85\linewidth]{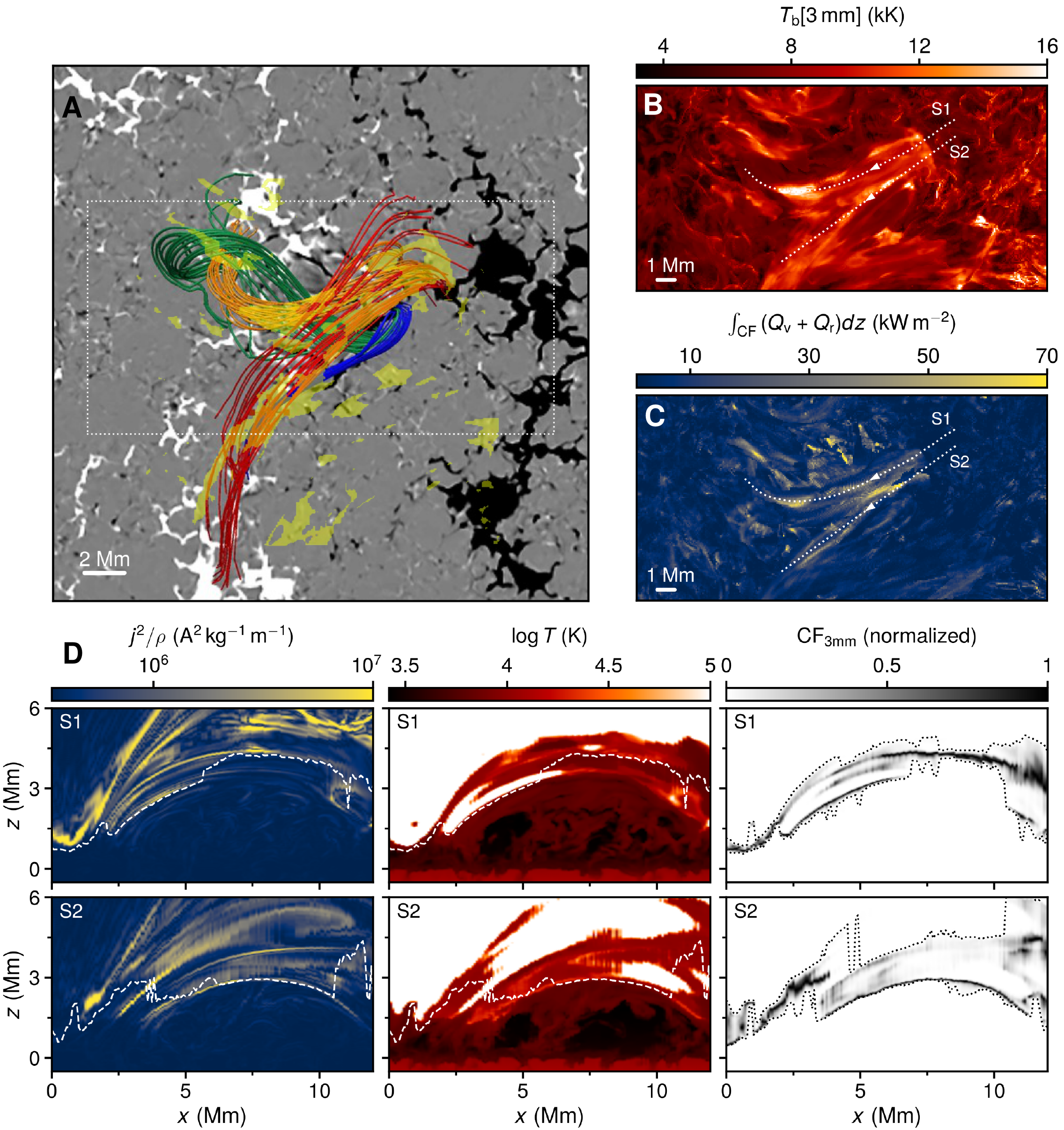}
    \caption{Magnetic topology, heating rates, and the formation of the millimeter continuum in a 3-D radiative-magnetohydrodynamics simulation. Panel A: Photospheric magnetogram clipped at +/- 1\,kG; the yellow shade shows where $T_{\rm b}[3 \mathrm{mm}]>9$\,kK. The area inside the dotted box is displayed in panels B and C, which show $T_{\rm b}[3 \mathrm{mm}]$ and the sum of viscous and resistive heating integrated within the CF of the 3\,mm continuum. The group of panels D show from the left to the right: Current density squared per mass unit (square-root scaling), logarithm of temperature, and CF$_{\rm 3\,mm}$ slices along the dotted paths {\it S1} and {\it S2} overlaid on panels B and C; the white dashed lines show where $\tau_{\rm 3\,mm}=1$, and the dotted black lines show where CF is at 1\% of the maximum in each column. } 
    \label{fig:fig4}
\end{figure*}

\section{Results}
\label{section:results}

AR 12738 showed ongoing magnetic flux emergence into a preexisting AFS. The ALMA map (Fig.\,\ref{fig:fig1}\textcolor{blue}{C}) reveals elongated patches of enhanced $T_{\rm b}\rm [3\,mm]$ by $\sim$3,000\,K relative to QS values, which is indicative of local heating in the upper chromosphere. The ALMA time series (supplementary movie) shows recurring brightenings exhibiting several hundred kelvin $T_{\rm b}\rm [3\,mm]$ variations, which last from a few tens of seconds to a few minutes with no clear periodicity. These hot spots are also fully visible in the AIA\,304\,\AA~images (Fig.\,\ref{fig:fig1}\textcolor{blue}{A}) and partially in AIA\,171\,\AA~(Fig.\,\ref{fig:Spl1}) -- but not in hotter channels. Correlation between Band\,3 brightenings and coronal emission in ARs has been reported before \citep{2020A&A...643A..41D,2021A&A...651A...6B}. 
There are other bright regions in the FOV in the 304\,\AA~images that have no counterpart in the ALMA maps. This may imply varying relative differences in formation heights of both diagnostics at different locations.

A comparison to the HMI LOS magnetogram (Fig.\,\ref{fig:fig1}\textcolor{blue}{D}) shows that this region is located between concentrations of magnetic field of opposite polarity, which are the footpoints of chromospheric loops that connect the two patches. The SST data indeed show short bright loop-like structure in the $\lambda$8542 core (Fig.\,\ref{fig:fig1}\textcolor{blue}{B}) and significant total linear polarization signals ($\mathrm{TLP}=\int\sqrt{Q^2+U^2}/I\,d\lambda$) indicative of strong transverse magnetic field in the photosphere and chromosphere (Figs.\,\ref{fig:fig1}\textcolor{blue}{E}, \ref{fig:fig1}\textcolor{blue}{F}). The composite image of other AIA EUV channels shows longer dark filaments overlying the small-scale loops (Fig.\,\ref{fig:Spl1}).

\subsection{Physical properties from inversions}

The non-LTE inversions provide a well-resolved temperature structure from the photosphere to the top of the chromosphere where the 3\,mm continuum is formed. The sensitivity of the spectra to magnetic fields is lower, and the inversions provide the vector magnetic field in the photosphere and mid-chromosphere. 

Figure~\ref{fig:fig2} shows the results of the \texttt{STiC} inversions. We find increased temperatures in the chromosphere at least up to optical depth $\log \tau$\,$\sim$\,$-6$ where $T$\,$\sim$\,10,000\,K, which is on the order of the observed $T_{\rm b}\rm\,[3\,mm]$. The warm locations feature $\lambda$8542 profiles with central reversals or raised intensities in the wings and small Doppler shifts ($\lesssim$\,$4\,\rm km\,s^{-1}$), which are well-reproduced by our models (supplementary Fig.\,\ref{fig:STiC_spl}). 

We computed the radiative energy losses as a proxy for the heating rate, which cannot be directly observed. In Fig.\,\ref{fig:fig2} we also show integrated losses from the inferred atmosphere in the strongest lines of H\,I, Ca\,II, and Mg\,II (Section\,\ref{section:enlosses}) in the height range spanned by the CF of the 3\,mm continuum. This is done in geometrical height scale assuming hydrostatic equilibrium.
Energy losses range from 2.6 to 4.9\,$\rm kW\,m^{-2}$ with a mean value of $\sim$4\,$\rm kW\,m^{-2}$ within the $T_{\rm b}\rm\,[3\,mm]=9$\,kK contours, which is higher than previous estimates of $\sim$2\,$\rm kW\,m^{-2}$ in the upper chromosphere 
\citep{1977ARA&A..15..363W}. 

The longitudinal and transverse photospheric field in the ROI have a maximum strength of $\lvert B_{\rm ln}\rvert$\,$=$\,$1890$\,G and $\lvert B_{\rm tr}\rvert$\,$=$\,$1380$\,G and the uncertainty is of the order of 10\,G. The chromospheric magnetic field is overall weaker, but the transverse component remains strong with an average (maximum) value of $\lvert B_{\rm tr}\rvert\sim480$\,G (1060\,G) with an uncertainty of $\sim$70\,G at $\log \tau=-5$ within the $T_{\rm b}\rm\,[3\,mm]=9$\,kK contours. The transverse field traces the near-horizontal tops of the emerging loops and coincides with higher temperature regions.

\subsection{Magnetic topology}

The fidelity of the magnetic field derived from the inversions is high, but the vertical resolution is low and the FOV is small. Therefore, we complement our analysis with a MHS extrapolation based on a high-resolution magnetogram derived from the SST $\lambda$6173 data (supplementary Fig.\,\ref{fig:ME}) embedded in a lower resolution magnetogram provided by HMI (Section\,\ref{section:extrapol}). 

Figure~\ref{fig:fig3} was produced using the \texttt{VAPOR} software\footnote{\url{www.vapor.ucar.edu}} \citep{atmos10090488} and it shows a 3D rendering of the overlying field connecting the pore and the plage region, whose direction is the same as the AFS in the EUV images (c.f. Fig.\,\ref{fig:Spl1}).

A comparison with Fig.\,\ref{fig:fig1}\textcolor{blue}{B} confirms that the field lines align with the direction of the bright $\lambda8542$ fibrils inside the ALMA $T_{\rm b}$ contours, in agreement with the picture drawn from the imaging and inversions. The horizontal magnetic field strength in the short loops derived from the extrapolation (as high as $\sim$\,800\,G around $z$\,$\sim$\,1\,Mm) is consistent with the inversion results. 
Patches of high $T_{\rm b}\rm\,[3\,mm]$ coincide with the interaction region between the loop systems where current sheets (tangential discontinuities) must exist between them. 

We computed the current density ($\vec{j}=\nabla \times \vec{B}/\mu_{0}$) from the extrapolation for completeness (supplementary Fig.\,\ref{fig:extrapolSpl}), but we note that spurious currents arise from the lack of smoothness in the magnetic field introduced by inversion noise in the magnetograms and the azimuth angle disambiguation, as well as limitations of the MHS extrapolation algorithm itself \citep{2018ApJ...866..130Z}. We find current strands over a range of heights relevant to the formation of the 3\,mm continuum ($\sim$1-4\,Mm, Fig.\,\ref{fig:fig4} and Fig.\,\ref{fig:simextra2}\textcolor{blue}{C}) that is cospatial with both the short $\lambda8542$ loops and the long, overlying ones seen in the AIA images, but we could not identify a single height that correlates strongly with $T_{\rm b}\rm[3\,mm]$. However, we do expect spatial and temporal variations of the formation height of the mm continuum \citep[e.g.,][]{2015A&A...575A..15L,2020ApJ...891L...8M}.

We attempted to correlate $T_{\rm b}[\rm 3\,mm]$ with $j^{2}/\rho$ in the MHS extrapolation at the height where $\tau_{\rm 3\,mm}=1$, obtained from the non-LTE inversions assuming hydrostatic equilibrium (typically $z$\,$\sim$\,1.5-1.8\,Mm) and taking into account the fact that the zero point of the MHS extrapolation is defined by the mean formation height of $\lambda6173$; however, this turned out to be inconclusive. This comparison was also made in column mass scale. 
This is partly due to the fact that the height scales obtained from these two methods are different, but we also expect the MHS extrapolation to underestimate the magnetic structure and field strength compared to $\lambda8542$ spectropolarimetry \citep{2021arXiv210902943V}, so the electric current values carry some uncertainty. Moreover, the height-integration effect of the CF of the 3\,mm continuum may smear such correlation, as further discussed in Section\,\ref{section:formation}.

Therefore, we cannot unambiguously link the observed millimeter emission to heating in current sheets based on these data alone. Nonetheless, the 3\,mm continuum forms above $\lambda8542$ \citep{2018A&A...620A.124D}, but below the He\,II\,304\,\AA~line, which strongly suggests that the mm continuum originates in the the shear layer. 

\subsection{Investigating the formation of the mm continuum}
\label{section:formation}

\begin{figure*}[t]
    \centering
    \sidecaption
    \includegraphics[width=120mm]{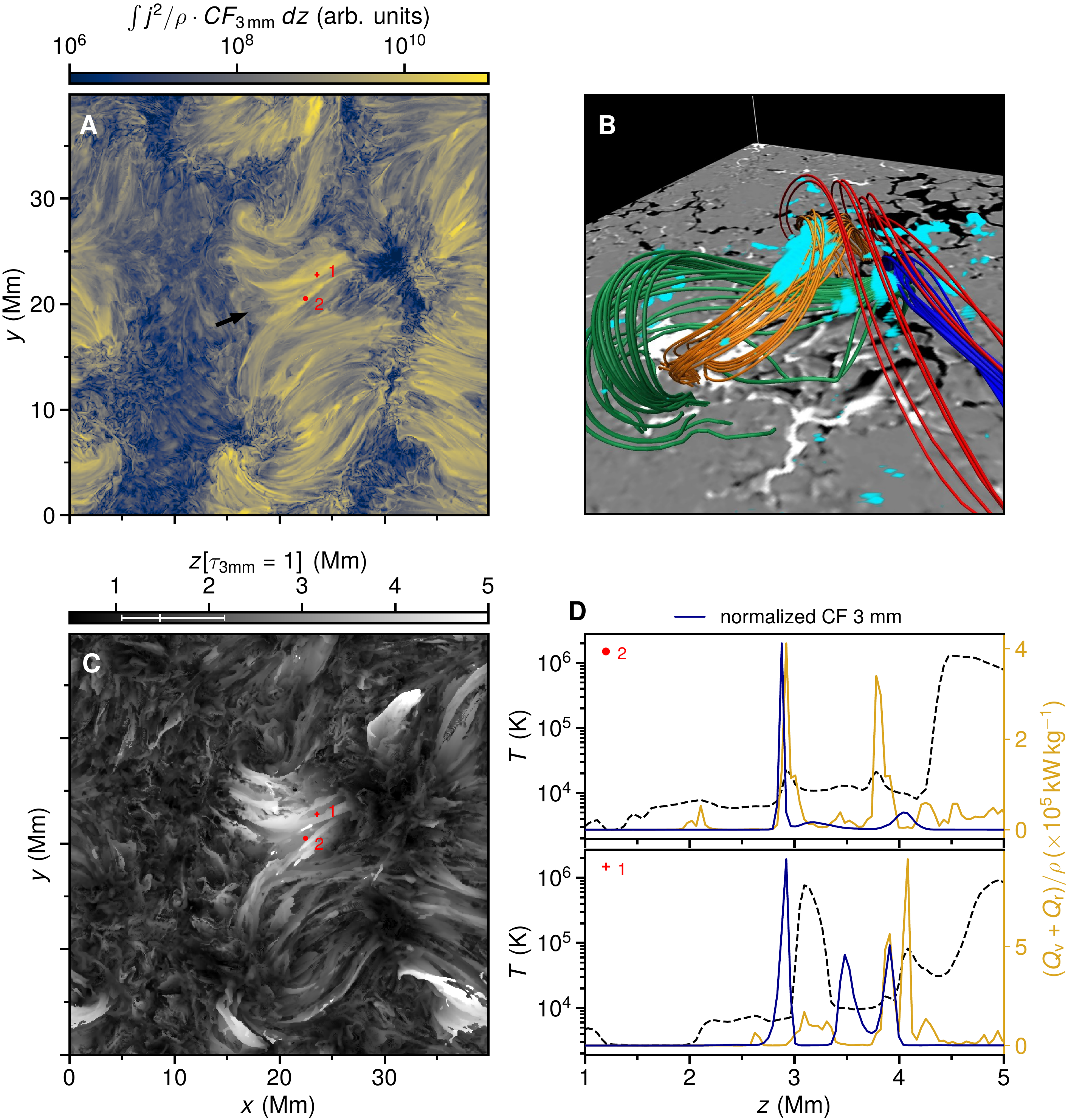}
    \caption{Heating rates and the formation of the millimeter continuum in the simulation. Panel A: Integrated $j^{2}/\rho$ weighted by the contribution function of the 3\,mm continuum (CF$_{\rm  3\,mm}$). Panel B: Photospheric magnetogram (range $\pm1$\,kG) with 3-D rendering of magnetic field lines and CF$_{\rm 3\,mm}$ (blue shade) of the region indicated by the arrow in panel A. A threshold was applied to the CF values not to hide the field lines. Panel C: Height at which $\tau=1$ at 3\,mm; the errorbar overlaid on the colormap shows the median and the range between the 16th and 84th percentiles of the distribution. Panel D: Temperature (dashed lines), total heating rates per mass (yellow lines), and CF$_{\rm 3\,mm}$ (blue lines) as a function of height at two locations indicated in panels A and C. }
    \label{fig:simextra2} 
\end{figure*}
\begin{figure*}
    \centering
    \sidecaption
    \includegraphics[width=120mm]{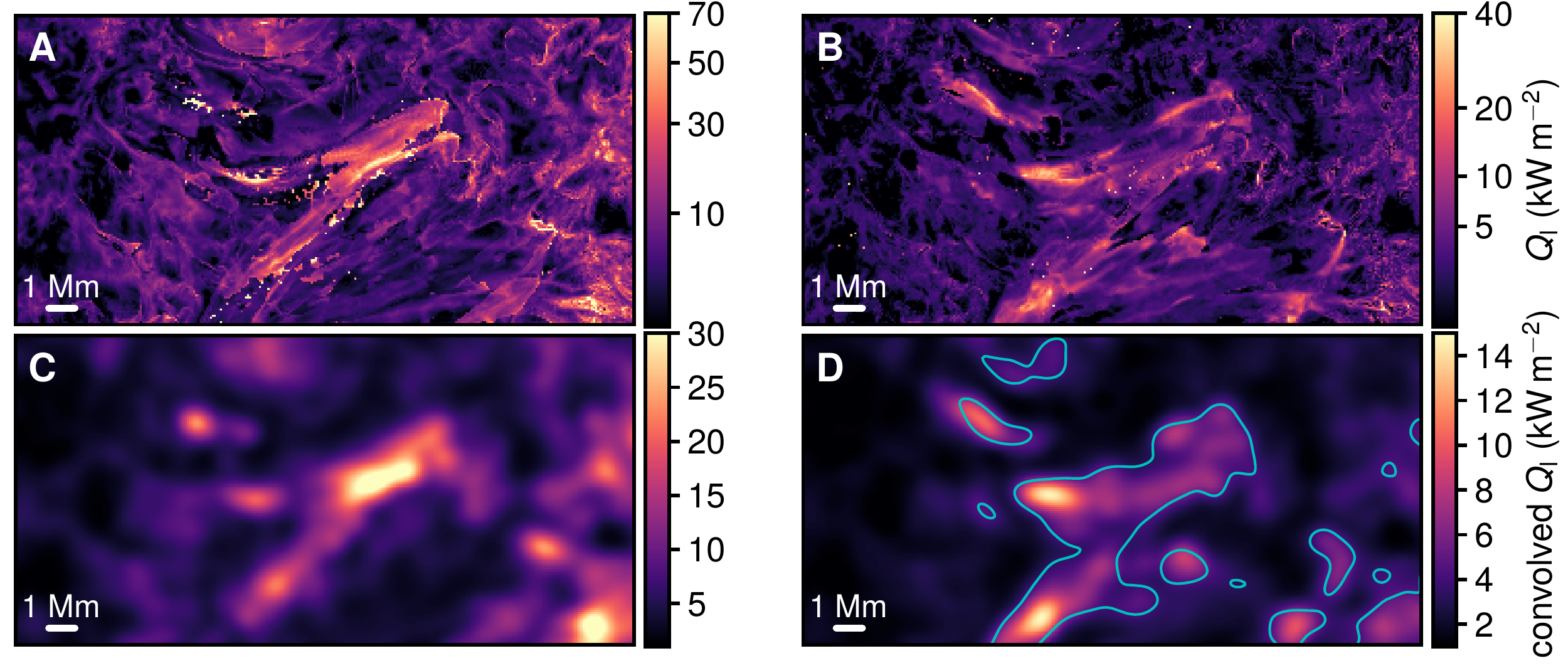}
    \caption{Radiative losses in the simulation. Total radiative losses integrated within CF$_{\rm 3 mm}$ from the \texttt{MURaM} output (panel A) and non-LTE values calculated using \texttt{STiC} (panel B) at full resolution; panels C and D show the corresponding quantities convolved with a Gaussian kernel with FWHM of 1.2$^{\prime\prime}$. Cyan contours show $T_{\rm b}[\rm 3\,mm]=9$\,kK. Top row panels are displayed in square-root scale.}
    \label{fig:simLosses}
\end{figure*}

To gain further insight into the magnetic flux emergence and how electric currents heat the atmosphere in that process, we ran a 3D r-MHD simulation using \texttt{MURaM}, as explained in Section\,\ref{section:simulation}. Figure~\ref{fig:simextra0} displays the strength of the vertical and horizontal components of the photospheric magnetic field vector along with the synthetic emission in the core of $\lambda8542$ (corrected for Doppler shifts) and $T_{\rm b}[\rm 3\,mm]$. In order to simulate the effect of the ALMA beam to first order, we convolved $T_{\rm b}[\rm 3\,mm]$ at the simulation resolution (Fig.\,\ref{fig:simextra0}\textcolor{blue}{D}) with a Gaussian kernel with full-width-at-half-maximum $\rm FWHM$\,$=$\,1.2$^{\prime\prime}$ (Fig.\,\ref{fig:simextra0}\textcolor{blue}{E}). In this paper we are primarily interested in the region inside the dashed box that encloses a patch of emerging flux and where we see some compact mm brightenings as well as bright fibrils. At the ALMA resolution the bright strands appear more blob-like as in the observations (Fig.\,\ref{fig:fig1}\textcolor{blue}{C}). The simulation is also able to reproduce $\lambda$8542 profiles qualitatively similar to the observations featuring central reversals and raised intensities in the wings. There are patches where the line core is in full emission, which we do not find in the single CRISP scan that we have.  Emission profiles are uncommon in the simulation, and they occur at sites of enhanced heating at the $\tau=1$ layer of the line core (e.g., panels S4 in Fig.\,\ref{fig:simLosses}). We would need a longer $\lambda$8542 time series to investigate whether these kinds of emission profiles can occur as in the simulation.

Figure~\ref{fig:fig4}\textcolor{blue}{A} shows the footprint of the emerging bubble together with field lines highlighting the direction of the emerging loops (green), the overlying filaments (red), and the two sets of reconnected field lines (blue and orange). The synthetic 3\,mm emission shows a bifurcated structure following the reconnected field lines in the interaction region 
(Fig.\,\ref{fig:fig4}\textcolor{blue}{B}). 

We find enhanced total heating rates, defined as the sum of the viscous, $Q_{\rm v}$, and resistive, $Q_{\rm r}$, heating within CF$_{\rm 3\,mm}$ in thin strands in the interaction region (Fig.\,\ref{fig:fig4}\textcolor{blue}{C}). The mean (maximum) values at the location where $T_{\rm b}\rm\,[3\,mm]>10,000$~K are 12\,kW\,m$^{-2}$ (184\,kW\,m$^{-2}$).
This heating is caused by dissipation of currents and viscous dissipation of mass flows, both of which are driven by the interaction of the magnetic loop systems. This does not only directly cause Joule heating, but also drives flows through the Lorentz force that dissipate through viscosity. Our simulation employs numerical diffusive and resistive terms with an effective magnetic Prandtl number $P_\mathrm{m} > 1$ so that viscous heating is the largest contributor. However, in the real chromosphere $P_\mathrm{m} < 1$ and the heating is dominated by electric resistivity. The 3\,mm continuum is optically thick where $P_\mathrm{m} < 1$ (supplementary Fig.\,\ref{fig:simSlices2}).

The vertical cuts along the slices {\it S1} and {\it S2} (Fig.\,\ref{fig:fig4}\textcolor{blue}{D}) underscore that CF$_{\rm 3\,mm}$ peaks at locations of high $j^2/\rho$ in the chromosphere where there are magnetic field gradients, but the resulting $T_{\rm b}\rm\,[3\,mm]$ may reflect contributions from multiple strands along the LOS. In supplementary Fig.\,\ref{fig:simSlices} we also provide additional 2D slices of the total heating rates per mass unit, which show the same qualitative picture. 
The CF$_{\rm 3\,mm}$ shows a loop-like structure following the shape of the chromosphere-transition region boundary where the gas is still partially ionized. Locations of larger $j^2/\rho$ at transition region and coronal temperatures have a negligible (3\,mm) opacity so they do not contribute to $T_{\rm b}\rm\,[3\,mm]$. However, the heating rate in those strands can be very large, and hot pockets embedded in the 3\,mm formation height range lead to the large peak values in Fig.\,\ref{fig:fig4}\textcolor{blue}{C}. 

The simulated chromosphere is pervaded by multiple current sheets at different heights and shows a complicated thermal structure (Figs.\,\ref{fig:fig4}\textcolor{blue}{D}), while the analysis of the opacity data shows that the formation height of the 3\,mm continuum varies significantly across the flux emergence region (Fig.\,\ref{fig:simextra2}\textcolor{blue}{C}). The CF of the 3\,mm continuum peaks where the loop systems meet an angle (Fig.\,\ref{fig:simextra2}\textcolor{blue}{B}) and the atmosphere is locally heated (see also the supplementary Fig.\,\ref{fig:simSlices}).  Figure~\,\ref{fig:simextra2}\textcolor{blue}{A} shows integrated $j^{2}/\rho$ weighted by CF$_{\rm 3\,mm}$, which reveals the filamentary structure of the heating that is very similar to the synthetic emission itself (c.f. Fig.\,\ref{fig:simextra0}\textcolor{blue}{D}).

The layer where optical depth is unity in the core of $\lambda$8542 is typically located below that of the 3\,mm continuum in the flux emergence region (supplementary Fig.\,\ref{fig:simSlices}). This suggests that the former traces the top of the low-lying fibrils, whereas the latter sees the heating at or above the $\lambda$8542 canopy. 

Figure~\,\ref{fig:simLosses} shows a comparison between the total radiative losses $Q_{\rm l}$ and $Q_{\rm l}[\rm STiC]$ in the simulated chromosphere. At the native resolution of the simulation, the losses show the same bifurcated structure in the interaction region (c.f. Fig.\,\ref{fig:fig4}\textcolor{blue}{B}, \ref{fig:fig4}\textcolor{blue}{C}). Total $Q_{\rm l}$ in the simulation reach above 100\,$\rm kW\,m^{-2}$ in some pixels due to the inclusion of losses at transition-region and coronal temperatures in thin pockets within the formation range of the millimeter continuum (c.f. the supplementary Fig.\,\ref{fig:simSlices}). These are not included in the \texttt{STiC} losses, hence, the lower values. Much of the filamentary structure is lost at the ALMA resolution, and the values of the radiative losses decrease significantly; the mean(standard deviation) is $\sim$\,$6(\pm 2)\,\rm kW\,m^{-2}$, where $T_{\rm b}\rm\,[3\,mm]$\,>\,9\,kK, which is similar to the values derived from the non-LTE inversions of the SST and ALMA data (Fig.\,\ref{fig:fig2}). 

\section{Discussion and conclusions}
\label{section:conclusions}

In this paper we present an analysis of joint SST and ALMA observations of an AR on the Sun using inversion methods, field extrapolations, and a numerical simulation. We find enhanced $T_{\rm b}[\rm 3\,mm]$ and bright $\lambda$8542 profiles between a pore and parasitic opposite polarity patch in the photosphere. This is also the location of short, low-lying chromospheric fibrils, where the magnetic field is more horizontal to the solar surface, underneath an overlying AFS seen in absorption in the AIA EUV channels.

Our analysis validates dissipation in current sheets as, at least, a locally dominant source of atmospheric heating, which produces brightenings in chromospheric diagnostics within ARs. Integrated radiative losses in the strongest chromospheric lines obtained from the simulation can be an order of magnitude higher than the ones determined from the observations. However, degrading the former to the spatial resolution of ALMA yields a mean(standard deviation) value of $\sim$\,6\,$(\pm 2)\,\rm kW\,m^{-2}$, which is consistent with the observed values. The heating occurs on spatial scales that are not resolved in the ALMA Band\,3 data. We note that this is only an approximate way of comparing losses obtained from observations and simulations at different spatial resolutions. In principle we would need to investigate the impact of the lack of constraints in the inversions on the estimates of the radiative losses by inverting synthetic data from the simulation at different resolutions, but this is beyond the scope of this paper.

Our estimates are much lower than the values reported in an observed magnetic reconnection event of up to $\sim$\,160$\,\rm kW\,m^{-2}$ \citep{2020arXiv201206229D}. This is partly due to differences in the integration method but, most importantly, that event was much stronger and showed strong flows and an associated surge, which we did not detect. However, our limited spectropolarimetric data only provide one time frame and do not allow us to investigate the dynamics in detail. The ALMA time series does show recurring elongated brightenings at that location prior to the SST campaign, which would be consistent with the bright strands that we see in the simulation.

We note that the simulation makes use of numerical diffusivity and viscosity terms that are larger than their physical values. This is common in numerical simulations of the type employed here, and experiments show that this has only a marginal effect on the total energy dissipation rate 
\citep{1996JGR...10113445G,2017ApJ...834...10R}, 
which is mainly determined by the Poynting flux input in the photosphere. 
The simulation was run with an effective numerical magnetic Prandtl number $P_\mathrm{m}>1$; about two thirds of the energy dissipated in the chromosphere is through viscosity and one third by electric resistivity. The chromosphere has a $P_\mathrm{m}$ that is significantly below unity, especially if ambipolar diffusion is taken into account 
\citep{2012ApJ...753..161M}, which appears as an additional cross-field resistivity in the MHD approximation. 
While the majority of the dissipation of magnetic energy in the simulated chromosphere occurs through the Lorentz force, which drives flows that are then damped through viscosity, the solar chromosphere causes the majority of the energy to dissipate through currents \citep[e.g.,][]{2017ApJ...834...10R,2019ApJ...879...57B}.

Although our simulation lacks some chromospheric physics such as ion-neutral interactions and non-equilibrium ionization \cite[e.g.,][]{2012ApJ...747...87K,2020A&A...633A..66N}, it reproduces well the observed magnetic configuration, infrared and radio spectra, and energetics remarkably. It also shows how the flux emergence leads to enhanced emission in the mm continuum, which is a good proxy for local chromospheric heating, and how the opacity may vary across the flux emergence region. However, the height-integration effect of the contribution function implies that the observed brightness temperatures may be a weighted average of contributions from several different layers along the LOS \cite[see also][]{2020ApJ...891L...8M}, which complicates the interpretation of observations.

The MHS extrapolation reveals a textbook magnetic topology similar to previously proposed models of the interaction of emergent flux and the canopy \cite[e.g.,][]{2003Natur.425..692S,2014LRSP...11....3C,2016ApJ...825...93O}, leading to the formation of current sheets between the two flux systems \cite[e.g.,][]{2007ApJ...666..516G,2014ApJ...788L...2A}. If the loop interaction occurs at coronal heights this may lead to much higher temperatures ($\gtrsim$\,1\,MK) and produce the recently discovered {\it campfire} EUV signatures, as simulations suggest \citep{2021A&A...656L...7C}.

These findings may play a role in explaining the solar cycle modulation of brightness temperatures in the millimeter range and their correlation with the sunspot number \citep{2020ApJ...902..136G}, as well as the excess millimeter brightness in chromospheres of other stars \citep{2015ApJ...809...47M,2015A&A...573L...4L,2017A&A...602L..10O}. Our analysis quantifies radiative losses, which can also be used to benchmark simulations of solar and stellar atmospheres. 

\begin{acknowledgements}

This paper makes use of the following ALMA data: ADS/JAO.ALMA\#2018.1.01518.S. The Swedish 1-m Solar Telescope is operated on the island of La Palma by the Institute for Solar Physics of Stockholm University in the Spanish Observatorio del Roque de los Muchachos of the Instituto de Astrofísica de Canarias. 
The Institute for Solar Physics is supported by a grant for research infrastructures of national importance from the Swedish Research Council (registration number 2017-00625). ALMA is a partnership of ESO (representing its member states), NSF (USA) and NINS (Japan), together with NRC (Canada), MOST and ASIAA (Taiwan), and KASI (Republic of Korea), in cooperation with the Republic of Chile. The Joint ALMA Observatory is operated by ESO, AUI/NRAO and NAOJ. 
Part of the calculations were performed on resources provided by the Swedish National Infrastructure for Computing (SNIC) at the National Supercomputer Centre (NSC) at Link\"{o}ping University and the PDC Centre for High Performance Computing (PDC-HPC) at the Royal Institute of Technology in Stockholm. 
This project has received funding from the European Research Council (ERC) under the European Union’s Horizon 2020 research and innovation program (SUNMAG grant agreement 759548 and SOLARNET grant agreement 824135), the Knut and Alice Wallenberg Foundation, Swedish Research Council (2021-05613) and Swedish National Space Agency (2021-00116); this material is partly based upon work supported by the National Center for Atmospheric Research, which is a major facility sponsored by the National Science Foundation under Cooperative Agreement No. 1852977; X.Z. is supported by the mobility program (M-0068) of the Sino-German Science Center. 

This research has made use of \texttt{Astropy} (\url{https://astropy.org}) -- a community-developed core Python package for Astronomy \citepads{astropy:2018} and \texttt{SunPy} (\url{https://sunpy.org}) -- an open-source and free community-developed solar data analysis Python package \citep{2015CS&D....8a4009S}.

\end{acknowledgements}

\bibliographystyle{aa} % style aa.bst
%\bibliography{bib.bib} % your references Yourfile.bib


\begin{appendix}

\section{Supplementary figures}

\begin{figure*}
    \centering
    \includegraphics[width=0.8\linewidth]{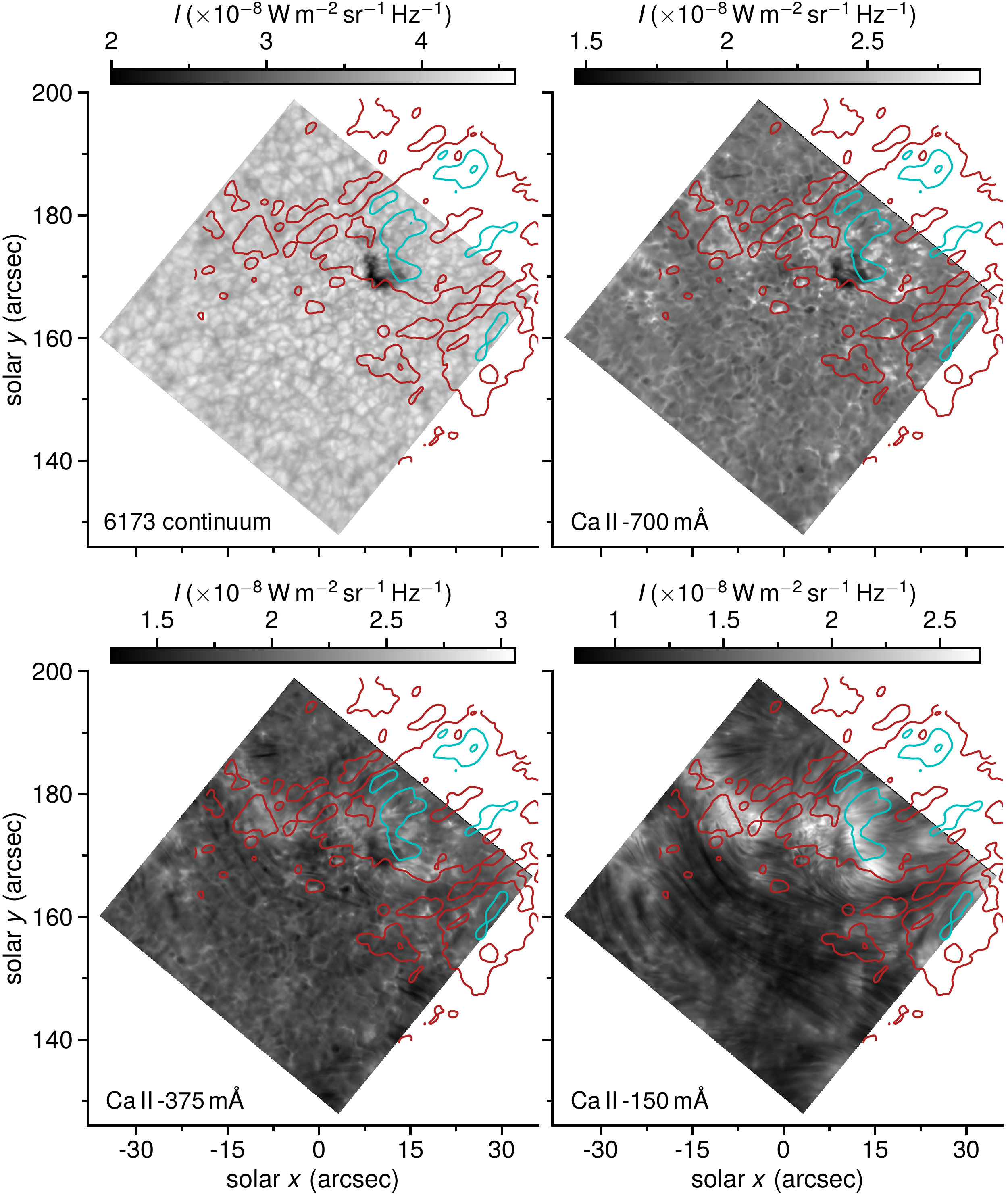}
    \caption{SST/CRISP photospheric and chromospheric filtergrams. Intensity in the continuum at 6173\,\AA~and at different wavelengths in the blue wing of $\lambda$8542. The red and cyan contours correspond to$T_{\rm b}\rm [3\,mm]= 8~and~9$\,kK mapped by ALMA.}
    \label{fig:Spl2}
\end{figure*}

Figure\,\ref{fig:Spl2} displays additional SST/CRISP context data, which clearly show the pore, surrounding plage, and overlying chromospheric fibrilar structures. Enhanced millimeter brightness (e.g., red contours) generally corresponds to the brightest regions of the $\lambda$8542 filtergrams.

\begin{figure*}
    \centering
    \includegraphics[width=0.8\linewidth]{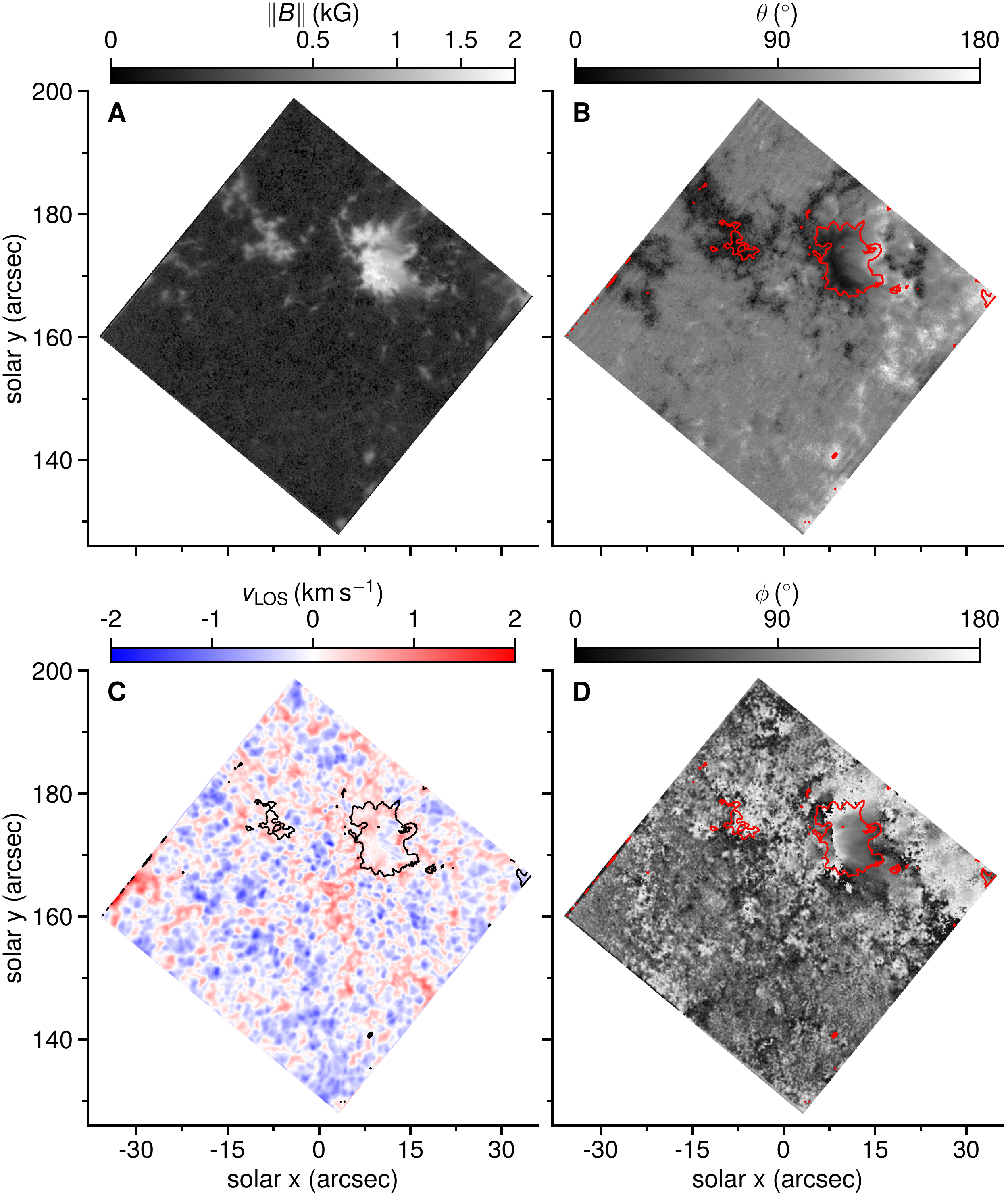}
    \caption{Milne-Eddington inversion of the SST/CRISP spectropolarimetry in $\lambda6173$. Panel A: Magnetic field strength with square-root colormap scaling; panel B: Inclination angle; panel C: Line-of-sight velocity; panel D: Azimuth angle. The contours correspond to $\Vert B\Vert=0.5$\,kG.}
    \label{fig:ME}
\end{figure*}

Figure\,\ref{fig:ME} shows the results of the ME inversion of the CRISP data. Residual fringe patterns in the Stokes V signals that could not be removed by either Fourier filtering or PCA leave a noticeable imprint on the inclination angle map, but mostly towards the NW and SE sides of the CRISP FOV. The fringing is not seen in the ROI and does not affect the results. The displayed azimuth angle map is not disambiguated but it is fairly smooth between the two opposite polarity patches. The LOS velocities show $\leq$\,1\,$\rm km\,s^{-1}$ upflows between the opposite photospheric polarities.

\begin{figure*}
    \centering
    \includegraphics[width=0.6\linewidth]{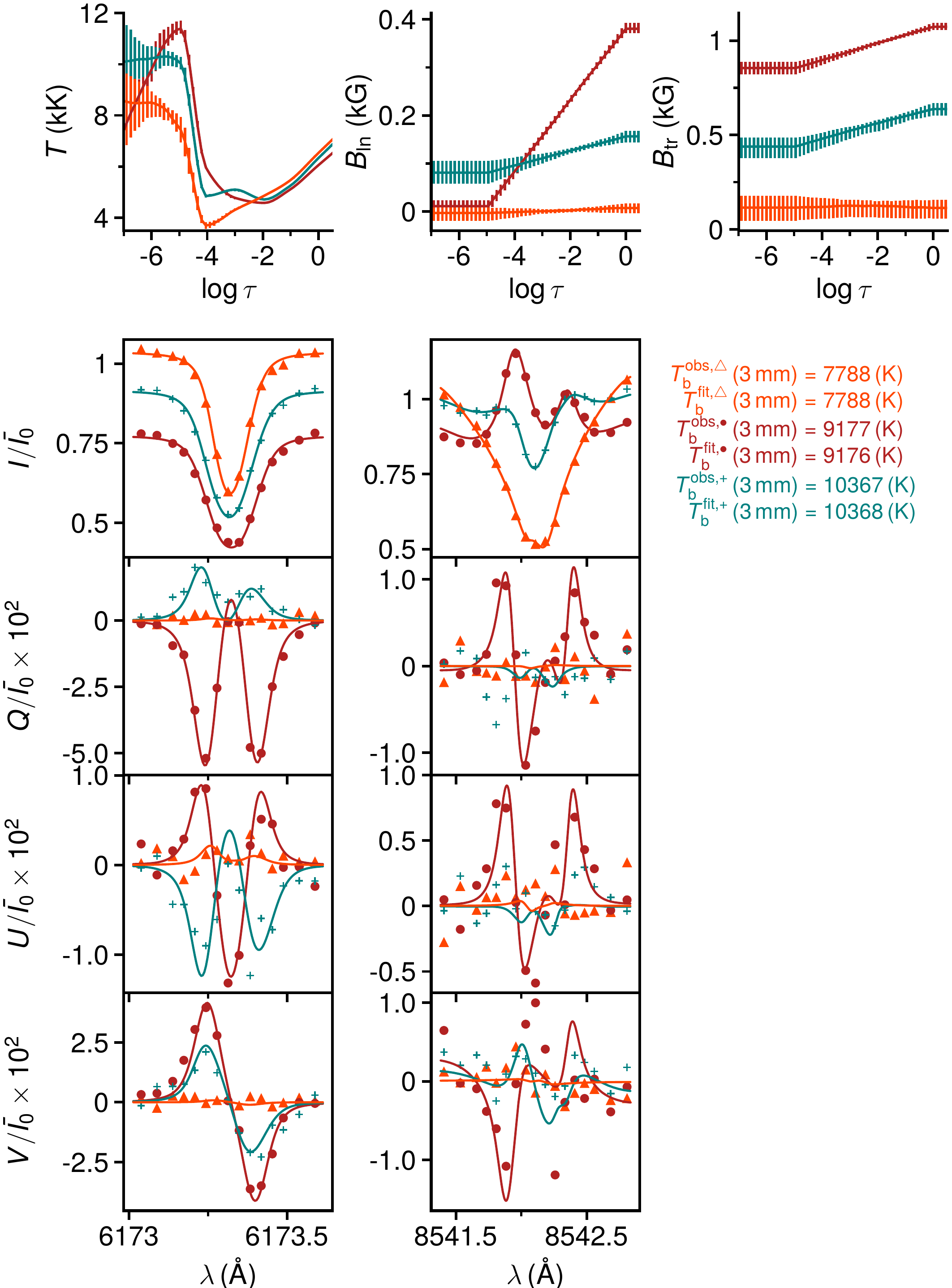}
    \caption{Observed and best-fit Stokes profiles and corresponding non-LTE models. The three example profiles (normalized by $\bar{I}_{0}$ -- the mean intensity in the "quiet" part of the FOV at the bluest wavelength) correspond to the markers shown in Fig.\,\ref{fig:fig2}. Observed and synthetic $T_{\rm b}\rm \,[3\,mm]$ values are indicated on the right. The solid lines and vertical bars in the top panels show the median and the range between the 16th and 84th percentiles of the Monte-Carlo distributions at each optical depth grid point. }
    \label{fig:STiC_spl}
\end{figure*}

Figure~\ref{fig:STiC_spl} shows the results of the Monte-Carlo \texttt{STiC} inversions on three selected pixels in the ROI. The spread in the physical parameters can be used to assess the uncertainties of the inverted models.

\begin{figure*}
    \centering
    \includegraphics[width=0.8\linewidth]{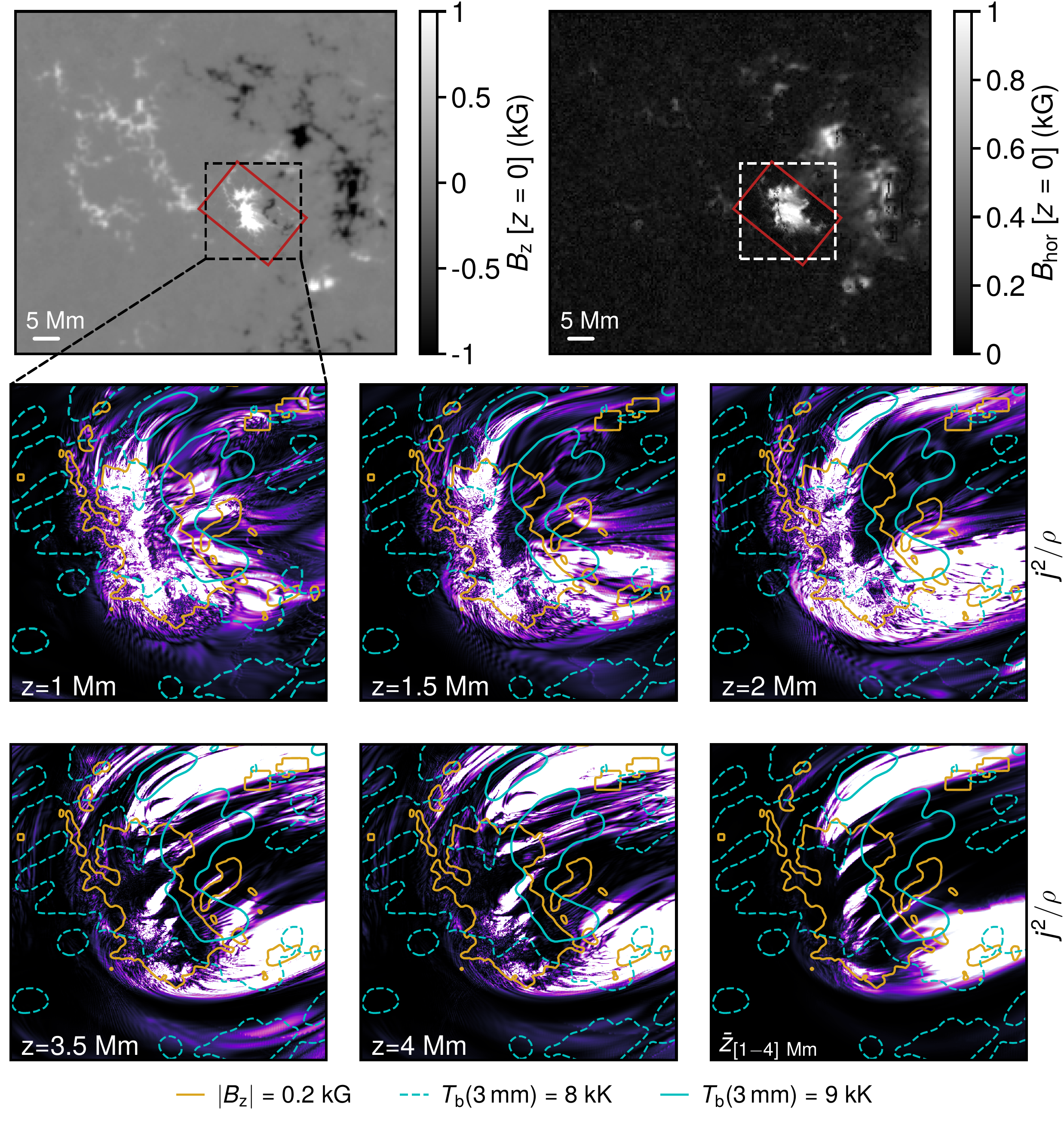}
    \caption{Current density calculated from the magnetic field extrapolation. The red rectangle in the top panels delimits the embedded SST/CRISP FOV in the extended HMI magnetogram. The middle and lower rows show $j^{2}/\rho$ at different heights in square-root scaling and arbitrary units for qualitative comparison with ALMA $T_{\rm b}$ contour overlays (cyan lines) within the region delimited by the dashed box. The lower right panel shows  $j^{2}/\rho$ averaged over height between $z=[1,4]$\,Mm. The yellow contour indicates $\lvert B_z \rvert$ at $z=0$\,Mm.}
    \label{fig:extrapolSpl}
\end{figure*}

Figure~\ref{fig:extrapolSpl} shows the photospheric composite vector magnetogram, which results from the combination of the SST and HMI data, along with the current densities at different heights calculated from the MHS extrapolation for qualitative comparison with the Band\,3 brightness contours. The range of heights was chosen based on the insight provided by the simulation.

\begin{figure*}[h]
    \centering
    \includegraphics[width=0.9\linewidth]{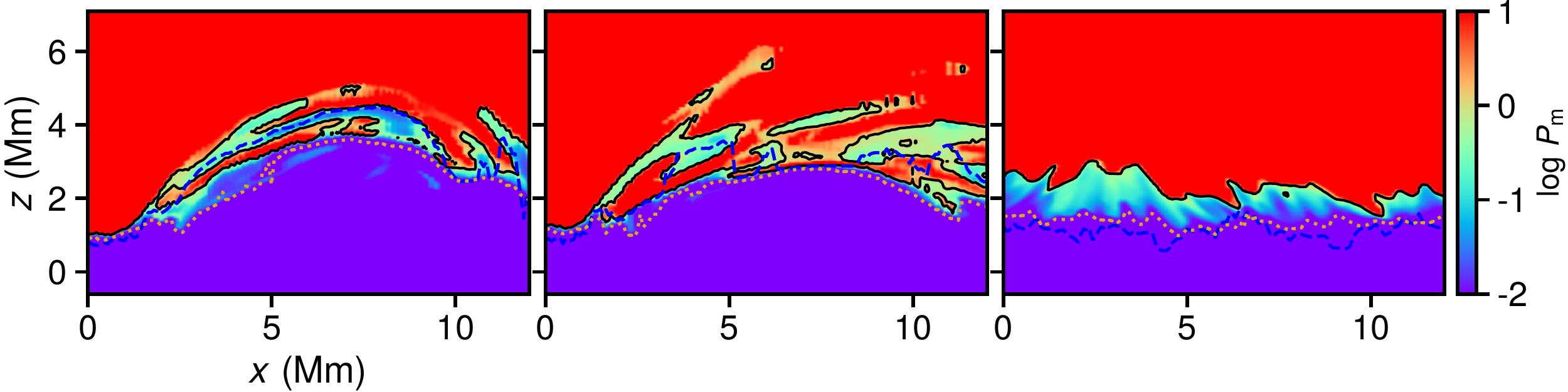}
    \caption{Magnetic Prandtl Number. Magnetic Prandtl number in the vertical cuts {\it S3}, {\it S4}, and {\it S5} (see supplementary Fig.\,10). The dashed line shows the $\tau=1$ layer of the 3\,mm continuum, the dotted line the same quantity for the core of $\lambda$8542. The solid black line shows $P_{\rm m}=1$. The colormap range is capped for display purposes.}
    \label{fig:simSlices2}
\end{figure*}

\begin{figure*}
    \centering
    \includegraphics[width=\linewidth]{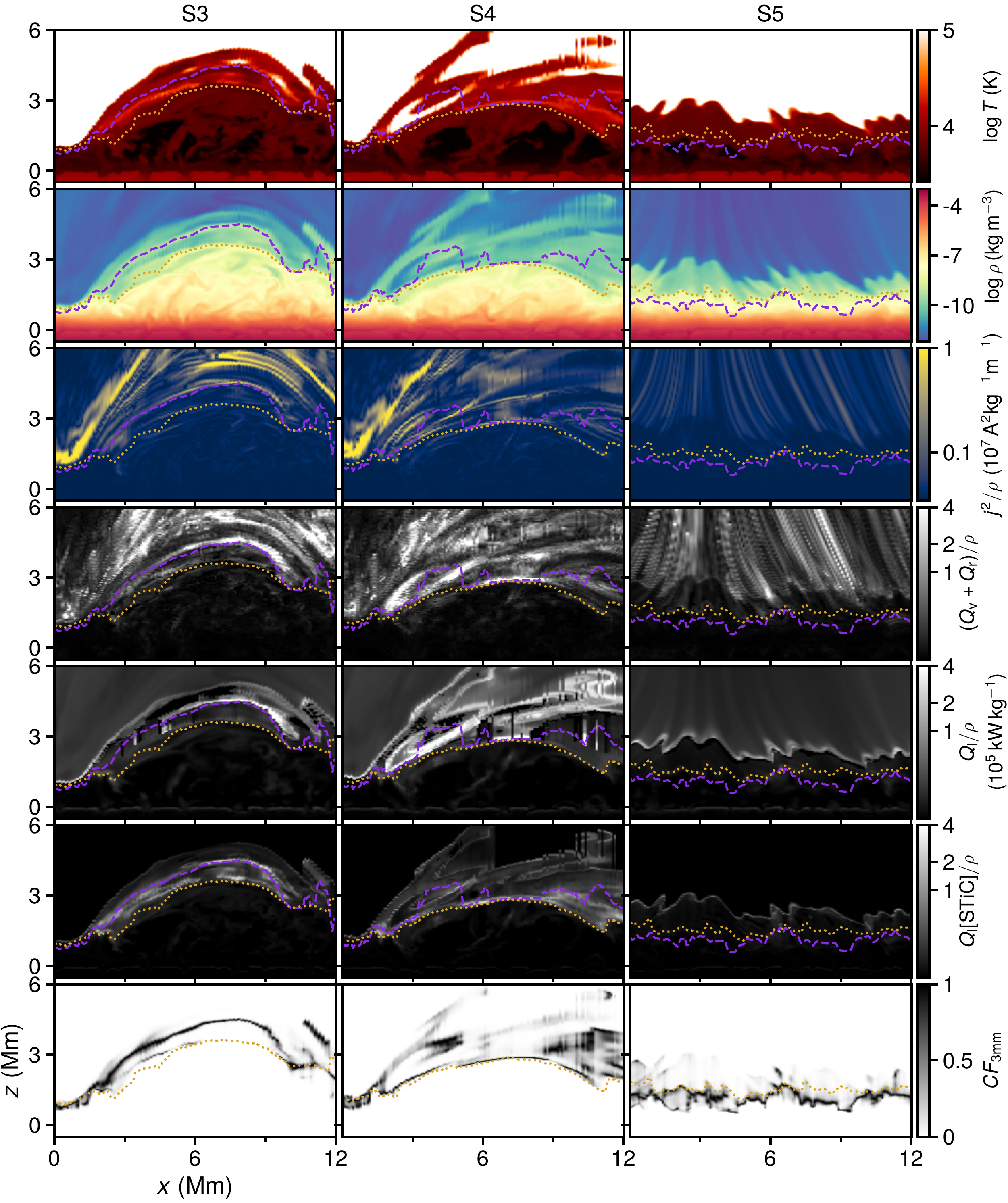}
    \caption{2D view of the simulation. Vertical cuts through the atmosphere along the slices {\it S3}, {\it S4}, and {\it S5} (see Fig.\,\ref{fig:simextra0}). The dashed line shows the $\tau=1$ layer of the 3\,mm continuum, the dotted line the same quantity for the core of $\lambda$8542. With the exception of the two top rows and the bottom row, all panels are displayed in power-law scale for display purposes.}
    \label{fig:simSlices}
\end{figure*}

Figure~\ref{fig:simSlices2} shows the magnetic Prandtl number -- the ratio of viscosity to the magnetic diffusivity, along 2D cuts through the simulated atmosphere (c.f. Fig.\,\ref{fig:simSlices}). Both $\lambda8542$ and the 3\,mm continuum are formed in the low $P_{\rm m}$ regime.

Figure~\ref{fig:simSlices} displays two vertical cuts through the flux emergence region  ({\it S3} and {\it S4}) and a cut through a part of the simulation with the open field ({\it S5}) for comparison. The formation height of the 3\,mm continuum follows layers of strong radiative cooling. The Joule heating proxy $j^2/\rho$ shows elongated current sheets, which are reflected in the heating terms ($Q_{\rm v}+Q_{\rm r}$). The radiative cooling, $Q_{\rm l}$, shows roughly the same structure, confirming our use of the radiative losses as a proxy for the heating. The quantity $Q_{\rm l}[{\rm STIC}]$ is lower because it only contains contributions from chromospheric lines.

\end{appendix}

\end{document}